\RequirePackage{ifpdf}
 \documentclass[preprint,11pt]{JHEP3}

\JHEPspecialurl{http://jhep.sissa.it/JOURNAL/JHEP3.tar.gz}

\usepackage{epsfig,multicol,amsmath}

\usepackage{epstopdf}
\usepackage{bm}  
\usepackage{amsmath}     
\usepackage{epsfig,epsf}
\usepackage{rotating}

\usepackage{cite}

\def\beq{\begin{equation}}
\def\eeq{\end{equation}}
\def\bea{\begin{eqnarray}}
\def\eea{\end{eqnarray}}

\def\eq#1{{Eq.~(\ref{#1})}}
\newcommand{\fig}[1]{Fig.~\ref{#1}}
\newcommand{\bas}{\bar{\alpha}_S}
\newcommand{\as}{\alpha_S}

\newcommand{\Lb}{\left(}
\newcommand{\Rb}{\right)}
\newcommand{\h}{\frac{1}{2}}
\parskip=\itemsep               

\newcommand{\nn}{\nonumber}

\newcommand{\ga}{\gamma}

\newcommand{\Ga}{\Gamma}
\newcommand{\om}{\omega}

\def\pom{{I\!\!P}}

\begin{boldmath}
\title{BFKL Pomeron with massive gluons}
\end{boldmath}
\author{\Large 
Eugene Levin,${}^{a,b}$ \,Lev  Lipatov${}^{c,d}$\, and\, Marat Siddikov${}^{a}$\\
 ${}^a$\, Departamento de F\'\i sica,
Universidad T$\acute{e}$cnica Federico Santa Mar\'\i a   and
Centro Cient\'\i fico-Tecnol$\acute{o}$gico de Valpara\'\i so,
Casilla 110-V,  Valparaiso, Chile\\
${}^b$ \, Department of Particle Physics, School of Physics and Astronomy,
Tel Aviv University, Tel Aviv, 69978, Israel\\
${}^c$\,Theoretical Physics Department,
Petersburg Nuclear Physics Institute,
Orlova Roscha, Gatchina,
188300, St. Petersburg, Russia\\
${}^d$\,Physics Department, St.Petersburg State University, Ulyanovskaya 3,
St.Petersburg 198504,
Russia}

\abstract
{
We solve the BFKL equation in the leading logarithmic approximation numerically in the Yang-Mills theory with the Higgs mechanism for the vector boson mass generation. It can be considered as a model
for the amplitude with the correct behavior of the $s$-channel partial waves at  large impact parameters. The Pomeron spectrum of
the massive BFKL kernel in the $\omega$-space for $t=0$ coincides with the continuous spectrum for the massless case although the density of its eigenvalues
is two times smaller for $\omega >\omega _0$, where $\omega_0$ is a negative number. We find a simple parametrization for the corresponding eigenfunctions.
Because the leading singularity in the $\omega$-plane in this Higgs model for $t=0$ is a fixed cut, the Regge pole contributions could be only for non-physical
positive $t$. Hence we can state that the correct behaviour at large $b$ does not influence the main properties of the BFKL equation.
}

\dedicated{PACS: 12.38-t, 12.38.Cy,1 2.38.Lg, 13.60.Hd, 24.85.+p, 25.30.Hm}
\preprint{TAUP 2975/14\\ USM-TH-320  \\
{\tt }\\
\today }

\begin{document}

\section{ Introduction}
The fundamental theoretical problem that has not been solved in the framework of CGC/saturation approach \cite{GLR,MUQI,MV,REV} is the large impact parameter ($b$) dependence of the scattering amplitude. As it has been discussed in Refs.\cite{KW1,KW2,KW3,FIIM}, the scattering amplitude at fixed $b$
in this approach satisfies the unitarity constraint being smaller than unity, but the radius of interaction increases as a power of energy leading to the violation of the Froissart bound\cite{FROI}.  Such power-like behaviour of the radius is a direct  consequence of the perturbative QCD technique which is a part of  the CGC/saturation approach. It stems from large impact parameter $b$ behaviour of the BFKL Pomeron\cite{BFKL,LIREV} which has the form: $A\Lb b \gg  1/Q_s\Rb \,\,\propto\,\,s^\Delta/b^2$. Amplitude $A\Lb b \gg  1/Q_s\Rb $ becomes of the order of unity at typical $b^2\, \propto\,\,s^{\Delta}$ leading to $\sigma\,\,\propto\,s^{\Delta}$ in the contradiction to the Froissart bound ($\sigma\,\,< c\,\ln^2 s$).  Since the lightest hadron (pion) has a finite mass ($m_\pi$) we know that the amplitude is proportional to $\exp\Lb - 2 m_\pi\,b\Rb$ at large $b$  instead of the power-like decrease. This exponential behaviour translates into the  Froissart bound. Therefore, we have to find how confinement of 
quarks and gluons 
being of non-perturbative nature, will change the large $b$ behaviour of the scattering amplitude.
 Since we are interested in the behaviour of the scattering amplitude at large $b$ where this amplitude is small, the non-linear effects can be neglected and one should introduce the non-perturbative corrections directly to the BFKL kernel.
 It has been checked by numerical calculations (see Refs.\cite{BEST1,BEST2,GBS1,GKLMN,LETA}) that if we modify the BFKL kernel introducing {\em by hand} a
 function that suppresses the production of the dipoles with sizes larger than $1/\mu_{soft}$, the resulting scattering amplitude has the exponential decrease at large impact parameters. 
 
 In this paper we are going to try a different way of modeling  the true large $b$ behaviour of the  BFKL kernel coming back to the  first papers on the BFKL Pomeron\cite{FKL}. In these papers it  is shown that the BFKL equation exists for  non-abelian gauge theories with the Higgs mechanism of mass generation. The kernel of the BFKL Pomeron, which depends on  the  Higgs mass,  falls down exponentially at large $b$ providing the finite
 radius of interaction that can grow only logarithmically and recovering the Froissart bound. Therefore, the BFKL equation with mass can be a training ground for answering the question: how the exponential  $b$-dependence at large $b$ could change the general features of the BFKL Pomeron and the CGC/saturation approach that is based on the BFKL equation. It should be stressed that the BFKL  Pomeron with the Higgs mass is closely related to the high energy asymptotic behaviour of the scattering amplitude in electroweak theory( see Ref. \cite{BLP}).
 
 In the next section we outline the derivation of the BFKL equation in the non-abelian theory with the Higgs mechanism of mass generation. This derivation was given in Ref.\cite{FKL} and we include it in the paper for the completeness  in order to present a coherent picture of the approach. In section 3 we discuss the main properties of the massive BFKL equation and prove that the maximum intercept of the massive BFKL Pomeron is  equal to the intercept  of the massless BFKL equation $ 4\bas \ln 2 $, where $\bas=N_c \alpha_S/\pi$. We find the numerical solution for the massive BFKL equation and give the simple approximate formulae both for eigenvalues and eigenfunctions of this equation. It turns out that for values $\omega \,\geq\,\omega_0\equiv -\h \bas$ the spectrum of the massive BFKL equation coincides with the spectrum of the massless BFKL equation. For momenta of gluons larger than mass,  the eigenfunctions approach the eigenfunctions of the massless BFKL equation while for momenta smaller than 
 mass, the eigenfunctions tend to be     constant values. For massive BFKL equation we detect that the eigenvalues in the vicinity of $\omega_0$ behave differently that for massless BFKL equation, and we propose the form of eigenfunctions that corresponds to this eigenvalue. In section 4 we investigate the energy behaviour of the average impact parameter for the massive BFKL. Generally speaking, such equation could generate the slope for the Pomeron trajectory since we introduce the dimensional parameter: mass. Solving equation we demonstrate that the massive BFKL equation leads to average impact parameter that is constant as a function of energy, repeating the behaviour of the massless BFKL equation.  In conclusion we discuss the main results of the paper.
 
 \section{Massive BFKL equation}
\begin{figure}[ht]
     \begin{center}
     \includegraphics[width=14cm]{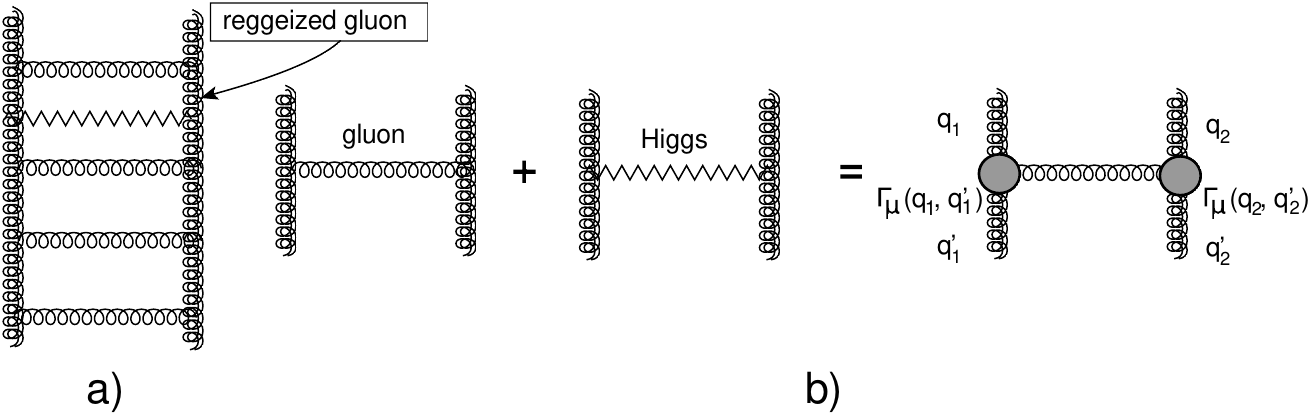} 
     \end{center}    
      \caption{ The massive BFKL equation (\protect\fig{eq}-a) and its kernel (\protect\fig{eq}-b)  }
\label{eq}
   \end{figure}

 The effective  vertex for the  gluon emission by the reggeized gluon in  the  Yang-Mills theory with the Higgs mechanism was calculated in Ref.\cite{FKL} and has a form (all notations are shown in \fig{eq})
 
 \beq \label{V}
 \Gamma_\mu\Lb q_1, q_1'\Rb\,\,=\,\,-  q^\perp_{1, \mu}\,-\,q'^\perp_{1, \mu} \,+\,p_{1, \mu}\Lb - \frac{q^2_1 + m^2}{p_1 \cdot k}\,+\,\frac{p_2\cdot k}{p_1\cdot p_2}\Rb\,-\,p_{2, \mu}\Lb - \frac{q'^2_
 1 + m^2}{p_2 \cdot k}\,+\,\frac{p_1\cdot k}{p_1\cdot p_2}\Rb, 
 \eeq
 where $q^2_i \,=\,|q^\perp_i|^2$ and $k_\mu\,=\,q_{1,\mu}\,-\,q'_{1,\mu}$ is the momentum of the emitted gluon.
 
 The gluon production vertex for the conjugated amplitude can be written as
 \beq \label{VC}
 \tilde{\Gamma}_\mu\Lb q_2, q_2'\Rb\,\,=\,\,-  q^\perp_{2, \mu}\,-\,q'^\perp_{2, \mu} \,+\,p_{1, \mu}\Lb - \frac{q^2_2 + m^2}{p_1 \cdot k}\,+\,\frac{p_2\cdot k}{p_1\cdot p_2}\Rb\,-\,p_{2, \mu}\Lb - \frac{q'^2_
 2 + m^2}{p_2 \cdot k}\,+\,\frac{p_1\cdot k}{p_1\cdot p_2}\Rb. 
 \eeq 
 Their product is equal to
 \beq \label{V2}
  \Gamma_\mu\Lb q_1, q_1'\Rb \cdot \tilde{\Gamma}_\mu\Lb q_2, q_2'\Rb \,=\,-2\Big(\frac{\Lb q^2_1 + m^2\Rb \Lb q'^2_2 + m^2\Rb}{k^2 + m^2}\,\,+\,\,\frac{\Lb q'^2_1 + m^2\Rb \Lb q^2_2 + m^2\Rb}{k^2 + m^2}\Big) \, +\,2 q^2\,+\,3 m^2,
   \eeq
   where $q_\mu \,=\,q_{1, \mu}\,-\,q_{2, \mu}\,=\,q'_{1, \mu}\,-\,q'_{2, \mu}   $.
   
   In the kernel of the BFKL equation ($K^{a' b'}_{a b}$), corresponding to the real particles in the intermediate state, this
product is multiplied by  $\as$ and by the corresponding color factor with an additional term
from the produced Higgs particles in the singlet and adjoint representations according to the model of  Ref.\cite{BFKL} (see \fig{eq}-b)
   \beq \label{KER} 
   K^{a' b'}_{a b}\,\,\propto\,\,\as\Big\{ - \h  \Gamma_\mu\Lb q_1, q_1'\Rb \cdot \tilde{\Gamma}_\mu\Lb q_2, q_2'\Rb f^{c a' b'} f_{c a b}\,\,+\,\,r\,m^2\Lb \delta_{a b}\,\delta^{a' b'}\,+\,   d^{c a' b'} d_{c a b}  \,\frac{N_c}{2}\Rb\Big\}  
 \eeq  
   where $f_{c a b}$  is the structure constant of the color group $SU(N_c)$, $d_{a b c}$ is the d-coupling
tensor and $\delta_{a b}$ is the Kronecker symbol. The coefficient $r$ can be fixed from
the bootstrap relation\cite{FKL}. Due to this relation in the adjoint representation
for the $ t$ -channel state the real contribution after its partial cancelation with the
virtual contribution, corresponding to the Regge trajectories, should be proportional
to $q^2 + m^2$.  Since  the projector on the adjoint representation is $ (1/N_c)f_{c a a'} f^{c b b'}$ we have
   \beq \label{BSTRP}
   K^{a' b'}_{a b}\frac{1}{N_c}\,f_{c a a'} f^{c b b'}   \,\,\longrightarrow\,\,- \h\Lb 2 q^2 + 3 m^2\Rb\frac{N_c}{2} \,+\,r m^2 
   \frac{N^2_c}{4}\,\,\sim\,\, -\left(q^2 + m^2\right)
   \eeq
   From \eq{BSTRP} we obtain
   \beq \label{r}
    r\,\,=\,\,\frac{1}{N_c}
    \eeq
    and the corresponding contribution to the kernel for the color singlet state in $t$-channel (BFKL Pomeron)
 is equal to   
    \bea \label{K0}
    K\Lb q_1, q_2 | q'_1, q'_2\Rb\,\,&=&\,\,\frac{\as}{2 \pi^2}\Big\{ - \h N_c   \Gamma_\mu\Lb q_1, q_1'\Rb \cdot \tilde{\Gamma}_\mu\Lb q_2, q_2'\Rb   \,\,+\,\,\frac{N^2_c}{4}\,m^2\Big\}\frac{1}{ (q'^2_1 + m^2) ( q'^2_2 + m^2)}\\
    &=& \frac{\as N_c}{2 \pi^2}\Big\{ \frac{1}{k^2 + m^2}\Big(\frac{q^2_1 + m^2}{q'^2_1 + m^2}\,+\,\frac{q^2_2 + m^2}{q'^2_2 + m^2}\Big)\,\,-\,\,\frac{q^2 +\frac{N^2_c+1}{N^2_c} m^2}{(q'^2_1 + m^2) ( q'^2_2 + m^2)}\Big\}\nn
    \eea   
    In the integral form the homogeneous BFKL equation at $ q = 0$ for the Yang-Mills theory with the Higgs mechanism is given by
    \beq \label{EQ}
\om f\Lb p\Rb\,\,=\,\,2 \om\Lb p \Rb f\Lb p\Rb\,+\, \frac{\as N_c}{2 \pi^2}\int d^2 p' \Big( \frac{ 2 f\Lb p'\Rb}{\Lb \vec{p}  - \vec{p}^{\,\,'}\Rb^2  +  m^2}\,\,-\,\, \frac{\frac{N^2_c+1}{N^2_c} m^2\,f\Lb p'\Rb}{(p^2 +m^2) (p'^2 + m^2}\Big)
\eeq
where we use the following notations: $ q_1 = q_2 = p$ and $q'_1 = q'_2 = p'$. 

The gluon Regge trajectory ($\om\Lb p \Rb$) is calculated explicitly, 
\bea \label{GTR}
\om\Lb |p| \Rb \,\,&=&\,\,- \frac{\as N_c}{4 \pi^2}\int \frac{ d^2 k \Lb p^2 + m^2\Rb}{\Lb k^2 + m^2\Rb \Lb\Lb  \vec{p} - \vec{k}\Rb^2 + m^2 \Rb}\nn\\
&=&\,\, - \frac{\as N_c}{2 \pi^2}\frac{|p|^2 +m^2}{|p|\sqrt{|p|^2 + 4  m^2}}\ln \frac{\sqrt{|p|^2 + 4 m^2} + |p|}{\sqrt{|p|^2 + 4 m^2} - |p|}
\eea
Assuming that we search the rotationally symmetric solution, the kernel can
be averaged over the azimuthal angle $\phi$
\bea\label{AZA}
\int^{2 \pi}_0 \frac{ d \phi}{2 \pi}\,\frac{1}{p^2 + p'^2 + m^2 - 2|p| |p'| \cos \phi}\,&=&\,\frac{1}{\sqrt{ \Lb p^2 + p'^2 + m^2\Rb^2 - 4 p^2 p'^2}}\nn\\
&=&\,\frac{1}{\sqrt{ \Lb p^2 - p'^2\Rb^2 \,+\,2 \Lb p^2 + p'^2\Rb m^2\,+\,m^4}}
\eea
Introducing the new variables\footnote{Besides variables $E$ and $\om$ we will use below the notation $\tilde{\om}= - E$ very often skipping tilde for simplicity. We hope that it will not lead to~misunderstanding since $\tilde{\om}$ is not proportional to $\bas$.}
\beq \label{VAR}
\kappa\,=\,\frac{p^2}{m^2};\,\,\,\,\,\,\,\,\,\,\,\,\,\,\kappa'\,=\,\frac{p'^2}{m^2};\,\,\,\,\,\,\,\,\,\,\,\,\,E\,=\,- \frac{\om}{\bas};\,\,\,\,\,\,\,\,\,\,\,\,\, \bas\,=\,\frac{\as N_c}{\pi}
\eeq
we obtain the one-dimensional BFKL equation
\bea \label{EQF}
\hspace{-0.5cm}&&E \phi\Lb \kappa\Rb\,\,=\\
\hspace{-0.5cm}&&\,\,\underbrace{\frac{\kappa +1}{\sqrt{\kappa}\sqrt{\kappa + 4  }}\ln \frac{\sqrt{\kappa + 4 } + \sqrt{\kappa}}{\sqrt{\kappa + 4 } - \sqrt{\kappa}} \phi\Lb \kappa\Rb}_{{\rm kinetic\, energy\, term}}\,\,-\,\,\underbrace{\int^{\infty}_{0}\,\frac{d \kappa' \phi\Lb \kappa'\Rb}{\sqrt{( \kappa - \kappa')^2\,+\,2 (\kappa + \kappa') + 1}}}_{\rm potential\, energy\, term}\,\,+\,\,\underbrace{\frac{N^2_c + 1}{2 N^2_c}\frac{1}{\kappa + 1}\int^{\infty}_0 \frac{\phi\Lb \kappa'\Rb \,d \kappa'}{\kappa' + 1}}_{{\rm contact\, term}}\nn
\eea

\section{Solution to the massive  BFKL equation}
\subsection{General features of the equation}
We start to discuss the solution to the equation considering the most general properties of solutions.  At
 large $\kappa$  solutions to this equation should coincide with the solution to the BFKL equation with $m = 0$ which has the following form:
\beq \label{EQBFKL}
E \,\phi_{\mbox{\tiny BFKL}}\Lb \kappa\Rb\,\,=\,\,\ln \kappa\, \phi_{\mbox{\tiny BFKL}}\,\,-\,\,\int^{\infty}_{0}\,\frac{d \kappa' \,\phi_{\mbox{\tiny BFKL}}\Lb \kappa'\Rb}{|\kappa - \kappa'|}
\eeq
after an appropriate regularization of divergency at $ \kappa' = \kappa $ (see \cite{BFKL})."

 The eigenvalues and the eigenfunctions of this equation are well known~\cite{BFKL,LIREV}. Therefore, the solution to \eq{EQF} has the following large $\kappa$ behaviour
\beq \label{LKAP}
\phi\Lb \kappa\Rb\,\,\xrightarrow{\kappa \to \infty}\,\,\phi_{\mbox{\tiny BFKL}}\Lb \kappa\Rb\,\sim\,\kappa^{-\h + i \nu}
\,\,\,\,\,\,\mbox{with}\,\,\,\,\,\,E\Lb \nu \Rb\,\,=\,\,\chi\Lb \nu\Rb\,\,=\,\,\psi\Lb \h + i\nu\Rb\,+\,\psi\Lb \h - i\nu\Rb \,-\,2 \psi\Lb 1 \Rb
\eeq
where $\psi(z) = d \ln \Gamma(z)/d z$ ( see formulae {\bf 8.36} in Ref.\cite{RY}).

Looking at  \eq{EQF} one can conclude that $\phi\Lb \kappa\Rb$ should be analytical functions with a cut at $\kappa < -4$ and pole at $\kappa=-1$.

We find instructive to re-write  \eq{EQ} in the coordinate representation. 

Using an identity
\beq \label{CORE1}
\int \frac{d^2 p'}{2 \pi} \frac{e^{i \vec{r} \cdot \vec{p}^{\,\,'}}}{ p'^2 + m^2}\,=\,\int^{+ \infty}_{-\infty}\frac{p' d p'\,J_0\Lb r p'\Rb}{p'^2 + m^2 }\,\,=\,\,K_0\Lb r m\Rb
\eeq
where $J_0\Lb z\Rb$ and $K_0\Lb z\Rb$ are the Bessel and Macdonald functions~\cite{RY}, 
we can rewrite \eq{EQ} in the form
\beq \label{H}
E\,f\Lb r \Rb\,\,=\,\,{\cal H}\,f\Lb r \Rb
\eeq
with
\beq\label{H1}
{\cal H}\,=\,\frac{p^2 +m^2}{|p|\sqrt{p^2 + 4  m^2}}\ln \frac{\sqrt{p^2 + 4 m^2} + |p|}{\sqrt{p^2 + 4 m^2} - |p|}\,\,-\,\,2 K_0\Lb |r| m \Rb\,+\,\frac{N^2_c + 1}{2\,N^2_c} \hat{P}\,\,=\,\,T\Lb p \Rb \,\,+\,\,V\Lb r \Rb \,\,+\,\,\frac{N^2_c + 1}{2\,N^2_c} \hat{P}\eeq
where $\hat{P}$ is  a shorthand notation for the projector onto the state $\sim m^2/(p^2+m^2)$
\beq \label{P}
\hat{P} \,\phi\Lb p \Rb\,=\,\frac{m^2}{p^2 + m^2}\int \frac{d^2 p'}{\pi} \frac{\phi\Lb p'\Rb}{p'^2\,+\,m^2}
\eeq
Let us introduce as a free Hamiltonian, the Hamiltonian for the massless BFKL equation (see \eq{EQBFKL}:
\beq \label{H0}
{\cal H}_0\,=\,\ln p^2 \,+\,\ln |r|^2\,-\,2 \psi\Lb 1 \Rb\,=\,\h\Lb \psi\Lb 1 + x \partial\Rb\, + \,\psi\Lb -x \partial\Rb\,+\, \psi\Lb 1 + x^* \partial^*\Rb\, + \,\psi\Lb -x^* \partial^*\Rb\,-\,4 \psi\Lb 1\Rb\Rb
\eeq

Since this Hamiltonian operates in the two-dimensional transverse plane, it is convenient to deal with the components  of all vectors as real and imaginary parts of the complex numbers, namely
\beq \label{COMPL}
x\,=\,r_1 + i r_2;~~~~~~ x^{*}\,=\,r_1 - i r_2; ;~~~~~~  \vec{p}\,=\,- i \vec{\nabla} \,=\,\Lb - i \partial - i \partial^{*},\,  \partial - \partial^{*}\Rb
\eeq
where the indices 1 and 2 denote the two transverse axes.

The eigenfunctions with the conformal spin $n = 0$  take the form (see Ref.\cite{LIREV})
\beq \label{EF}
f^{\pm \nu}_0\Lb |r|\Rb\,\,=\,\,|r|^{-1 \pm 2 i \nu}
\eeq
with the eigenvalues  $E\Lb \nu \Rb$ given by \eq{LKAP}.
The eigenfunctions of \eq{EF} have the following orthogonality and completeness properties
\bea 
\int^\infty_0 d |r|^2 f^{ \nu}_0\Lb |r|\Rb f^{ \mu *}_0\Lb |r|\Rb\,\,&=& \,\,2 \pi \delta\Lb \mu - \nu\Rb\,; \label{ORTCOM1}\\
|r| |r'| \int^{+ \infty}_{-\infty} d \nu f^{ \nu}_0\Lb |r|\Rb f^{ \nu *}_0\Lb |r'|\Rb\,\,&=& \,\,2 \pi \delta\Lb \ln
|r|^2\,-\,\ln|r'|^2\Rb\,.\label{ORTCOM2}
 \eea

The Green function for the free Hamiltonian  satisfies the following equation
\beq \label{GFH0}
\Lb E \,-\,{\cal H}_0 \Rb G_0\Lb r, r'\Rb\,\,=\,\,\frac{2 \pi}{|r| |r'| }\delta\Lb \ln
|r|^2\,-\,\ln|r'|^2\Rb
\eeq and it has the form
\beq \label{GFH01}
G_0\Lb r, r'\Rb\,\,=\,\,\frac{1}{ |r| |r'|}\int^{+ \infty}_{-\infty}\frac{d \nu}{E \,-\,E\Lb \nu\Rb } \Lb \frac{|r|}{|r'|}\Rb^{ 2 i \nu}
\eeq
The Green function for the general Hamiltonian of \eq{H} can be found as a solution to the integral equation
\beq \label{GFH}
G\Lb r, r'\Rb\,\,=\,\,G_0\Lb r, r'\Rb\,\,+\,\,\int d r'' G_0\Lb r, r''\Rb\,\,\Big( {\cal H}\,-\,{\cal H}_0 \Big) G\Lb r'', r'\Rb
\eeq
\eq{GFH} gives a natural way for applying a perturbative approach. In particular, in the lowest order of expansion with respect to $m^2$ we have
\beq \label{LOM}
{\cal H}\,-\,{\cal H}_0\,\,=\,\,\frac{m^2}{p^2}\Big( - \ln \frac{p^2}{m^2} + 2 \Big) \,+\,\frac{m^2 r^2}{4} \Big( \ln \frac{r^2}{4}  - 2 \psi\Lb 2 \Rb\Big) \,+\,\frac{N^2_c + 1}{2\,N^2_c} \hat{P}\,\,+\,\,{\cal O}\Lb m^4\Rb
\eeq

At large distances ($ r \to \infty$) the potential energy in Hamiltonian $\Lb V\Lb r \Rb = - 2 K_0\Lb r m \Rb\Rb$ is exponentially small, the contribution from the projector $\hat{P}$ in \eq{H} is proportional to $1/(p^2 + m^2)$ and is also exponentially suppressed, so the only relevant term in the hamiltonian is the kinetic energy
\beq \label{KINE}
 E\Lb p\Rb\,\,=\,\,T\Lb p \Rb\,\,=\,\,\frac{p^2 +m^2}{|p|\sqrt{p^2 + 4  m^2}}\ln \frac{\sqrt{p^2 + 4 m^2} + |p|}{\sqrt{p^2 + 4 m^2} - |p|},
 \eeq
 for which the eigenfunctions have a form
\beq \label{PLWA}
 f\Lb \vec{r}\Rb\,\,\sim\,\, e^{ i \sqrt{p^2} r}, p^2>0,\quad f\Lb \vec{r}\Rb\,\,\sim\,\, e^{ - \sqrt{-p^2} r}, p^2<0.
 \eeq
The point $p=0$ is special since it separates two different behaviours at large  $r$. This point corresponds to energy $E=\h$ or $\omega=\omega_0\equiv -\h\bas$. As we will see below, there are qualitative changes in the shape of the wave functions near this point. From the structure of the kinetic energy term~(\ref{KINE}) we can see that the energy $E$ is positive ($\omega<0$) for $p^2>0$, however for $-4m^2<p^2<0$ the energy may  have any value from $-\infty$ up to $\h$. In reality the spectrum $E$ is limited from below by $-4\ln 2$, as it is shown in sections 3.2 and 3.3.

 In the small-$r$ limit the eigenfunctions should approach the eigenfunctions of the massless BFKL equations, $ f^{\pm \nu}_0\Lb |r|\Rb$~\eq{EF}, with the spectrum given by~\eq{LKAP}. 
 
 Combining~\eq{EF} and~\eq{KINE}, we may get the relation between the parameters $\nu$ and $p$, which control the the small-$r$ and large-$r$ asymptotic behaviour, 
 
     \begin{figure}[ht]
     \begin{center}
     \includegraphics[width=10cm]{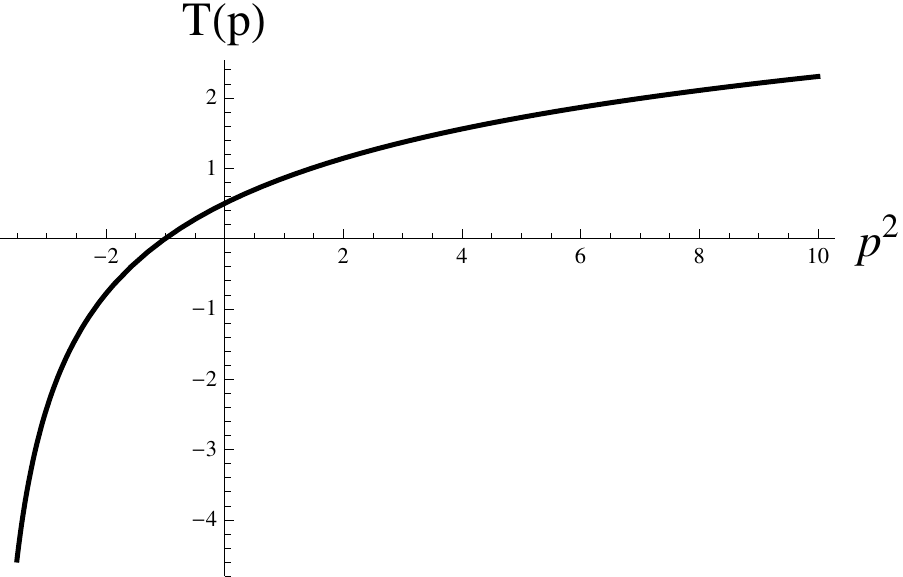} 
     \end{center}    
      \caption{ The dependence of the kinetic energy (see \protect\eq{KINE}) versus $p^2$ for $m =1$. }
\label{t}
   \end{figure}

 \beq \label{ABEQ}
 E=T(p)=\chi(\nu)
 \eeq

 \subsection{Estimates from the variational method}
 In the variational approach the upper bound for the ground state energy $E_{0}$ of the hamiltonian $\mathcal{H}$ may be found minimizing the functional
 \beq \label{VP1}
 E_{\mbox{ground}}\,\,\equiv\,\,E_0\,\,\,\leq \,\,\,
 F[\{\phi\}]\,\,=\,\,\frac{\Big{\langle} \phi^{\ast}\!\Lb r \Rb \Big{|} {\cal H}\Big{|} \phi\!\Lb r\Rb \Big{\rangle}}{\Big{\langle} \phi^{\ast}\!\Lb r \Rb \Big{|} \phi\!\Lb r\Rb \Big{\rangle} }
 \eeq
  \eq{VP1} means that the functional $F[\{\phi\}]$ has a minimum for function $\phi_0\Lb r \Rb$ which is the eigenfunction of the ground state with energy $E_0$.

 For our Hamiltonian in the momentum space \eq{VP1} can be re-written in the form
\beq\label{VP2}
 E_{0}\,\,=\,\,\min_{\phi}\frac{\int^\infty_{0} d \kappa \, T\Lb \kappa\Rb \, | \phi\Lb \kappa\Rb|^2  \,-\,\int^\infty_0 d \kappa \int^\infty_0  d \kappa' \frac{\phi\Lb \kappa\Rb \phi^*\Lb \kappa'\Rb}{\sqrt{\Lb\kappa - \kappa'\Rb^2 + 2 \Lb \kappa + \kappa'\Rb + 1}}\,+\,\frac{N^2_c + 1}{2 N^2_c} \Big{|} \int^\infty_0  d \kappa\, \frac{\phi\Lb \kappa\Rb }{\kappa + 1}\Big{|}^2}{\int^\infty_0 d \kappa  | \phi\Lb \kappa\Rb|^2}
\eeq

 The success of finding the value of $E_0$ depends on the choice of the trial functions in \eq{VP2}.  We choose it in the form
 \beq \label{TRIF}
 \phi_{\mbox{trial}}\Lb \kappa\Rb\,\,=\,\,\frac{1}{\Lb \kappa + a^2\Rb^{\ga}}
 \eeq 
 In the coordinate representation~\eq{TRIF} corresponds to
 \beq \label{TRIFCR}
 f_{\mbox{trial}}\Lb r \Rb\,\,=\,\,\frac{1}{\Ga\Lb \ga\Rb} \Lb \frac{r}{ 2 a}\Rb^{- 1 + \ga}\,K_{1 - \ga}\Lb a r \Rb \,\,\rightarrow\,\,\left\{ \begin{array}{l l} r \,\rightarrow \infty\,\, & \,\,\, \propto ~~r^{-\ga} e^{- a |r|}\\
  &\\
  r \,\rightarrow\,\, 0 &\,\,\, \propto\,~~r^{-2 + 2 \ga} \end{array}\right.
  \eeq
 One can see that our trial function has the correct behaviour if $a >0$ and $ b\,=\,2 \ga - 1\,>\,0$.
     \begin{figure}[ht]
    \begin{tabular}{c c c}
      \includegraphics[width=7.5cm]{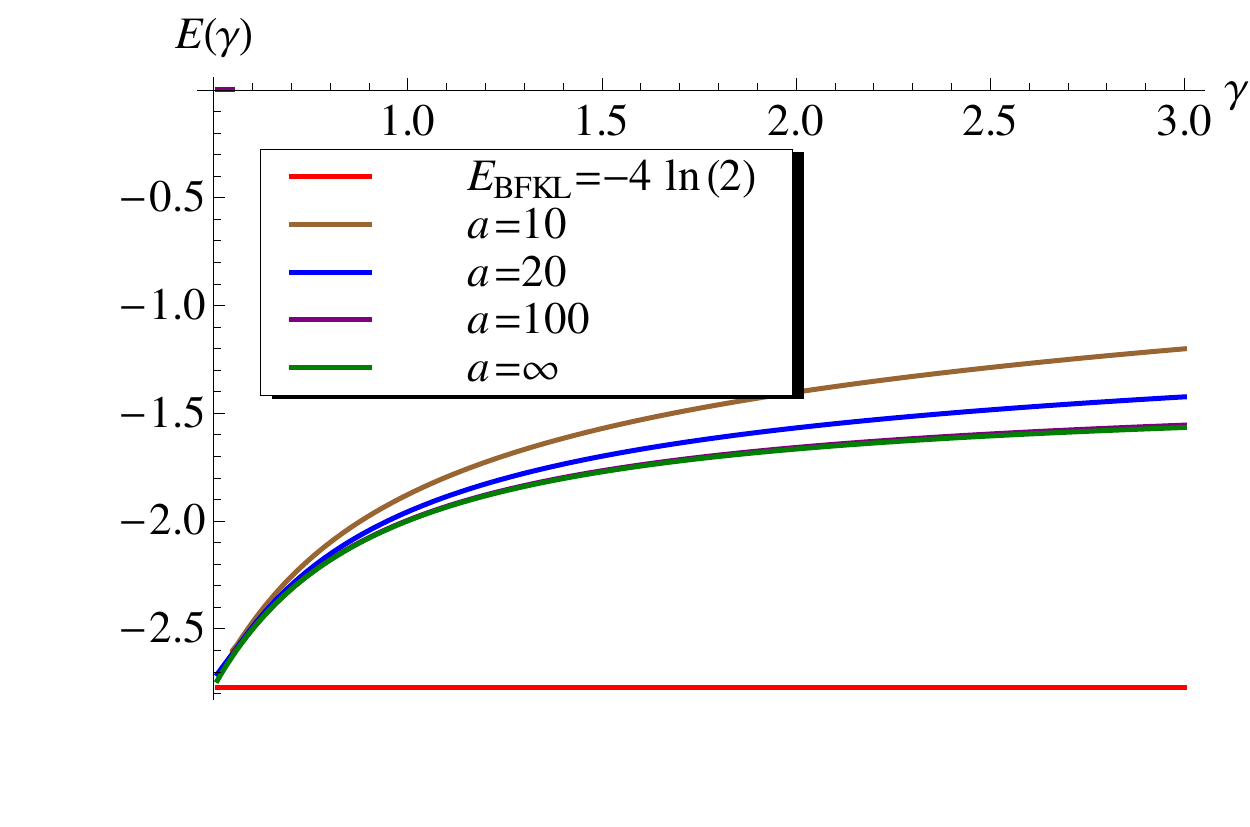} & ~~~~~~~& \includegraphics[width=7cm]{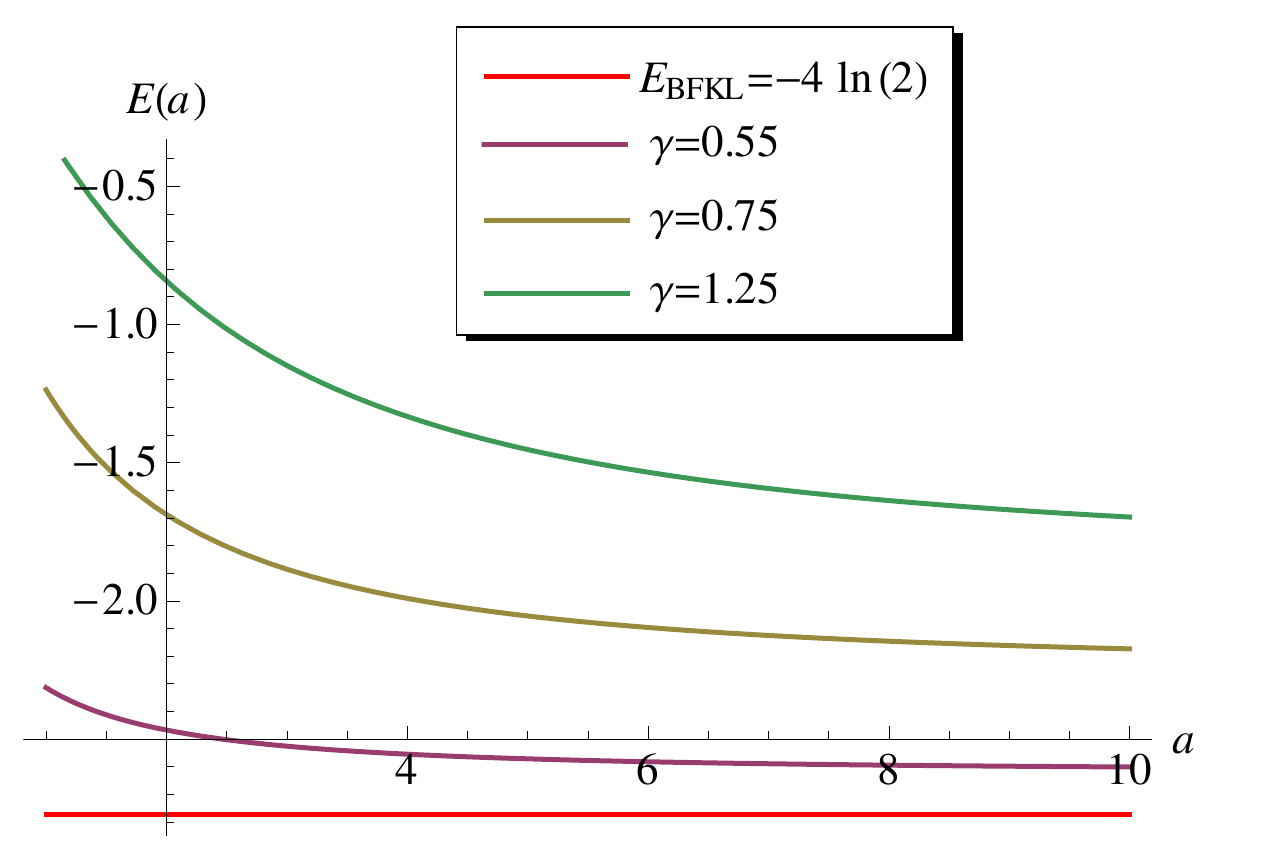} \\
      \fig{var}-a & &\fig{var}-b\\
      \end{tabular}
      \caption{ Dependence of $E_0$ given by \protect\eq{VP2} on $\ga$ (see \protect\fig{var}-a) and $a$(see \protect\fig{var}-b). The red straight line corresponds to the ground state energy of the massless BFKL equation $E_{ \mbox{\tiny BFKL}} \,\,=\,\,- 4 \ln 2$.
      }
\label{var}
   \end{figure}

     \begin{figure}[ht]
     \begin{center}
     \includegraphics[width=10cm]{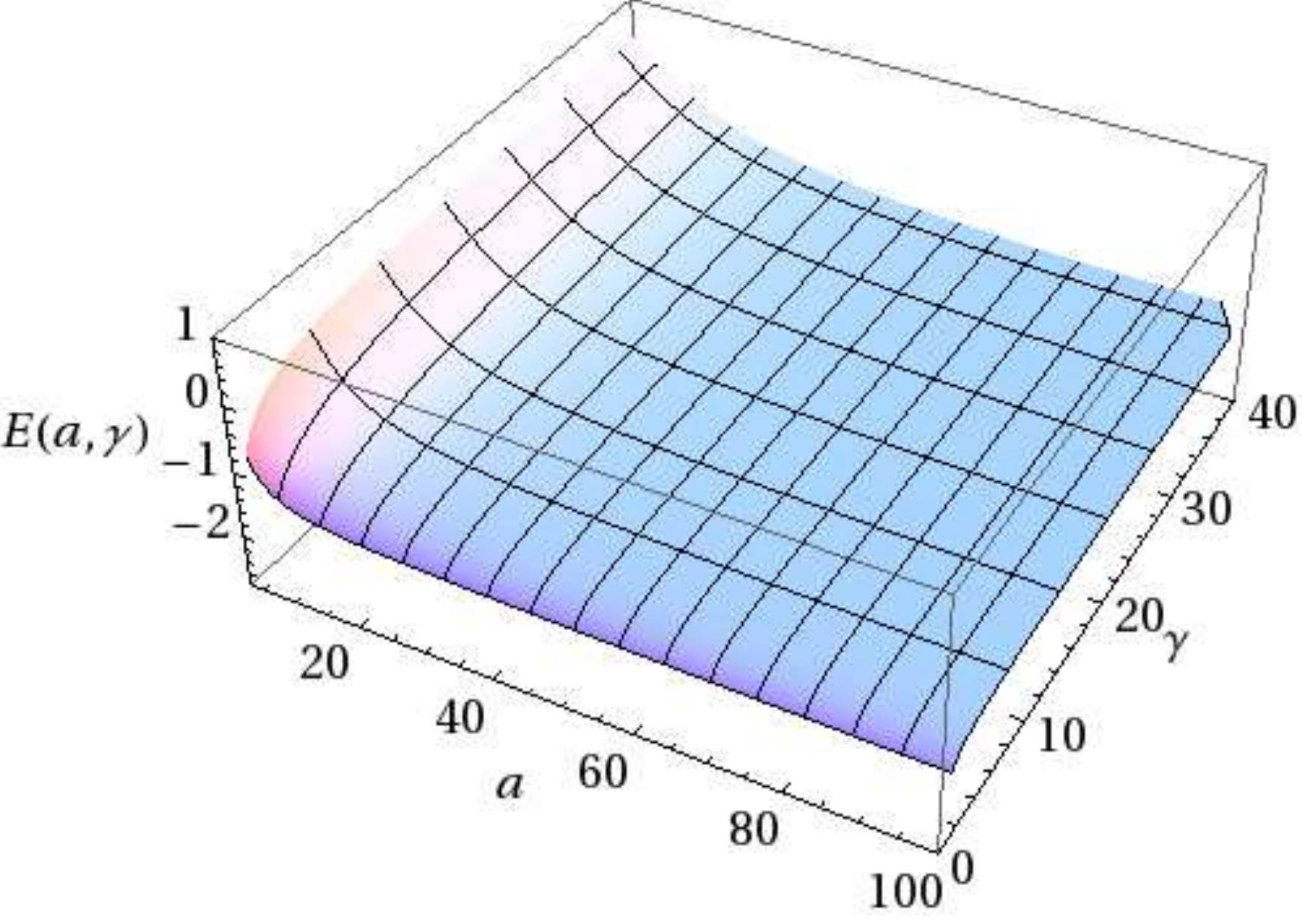} 
     \end{center}    
      \caption{ The dependence of  the energy given by \protect\eq{VP2} on the values of parameters $a$ and $\ga$. }
\label{var3d}
   \end{figure}

 \fig{var} shows the dependence of $E_0$ on  $\ga$ and $a$. At large $a$ and $\ga \to 0.5$, $E_0$ reaches minimum value which is the massless BFKL  energy $E_{BFKL}$. Therefore, we  conclude that the ground state energy $E_0$ could be only smaller than $E_{BFKL}$ but not larger than it. \fig{var3d} demonstrates the global tendency in the dependence of $E_0$ on the values of parameters $a$ and $\ga$.
 Similar results were obtained for more complicated parameterizations
like 
\begin{equation}
\phi(\kappa)=\frac{\kappa^{-\delta}}{\left(\kappa+a^{2}\right)^{\gamma}},
\end{equation}

\begin{equation}
\phi(\kappa)=\frac{\kappa^{-\delta}}{\left(\kappa+a^{2}\right)^{3/2}\left(\kappa+b^{2}\right)^{\gamma-3/2}}.
\end{equation}


While from the variational principle we always obtained the energy $E>E_{BFKL}$, we believe that the true minimum of 
the energy is $E= E_{BFKL}$ (respectively the eigenvalue $\omega=\omega_{BFKL}$), there is no indication that there are eigenvalues with $\omega>\omega_{BFKL}$. Actually,  with trial function of \eq{TRIF} for $a \gg 1$ we can perform the analytical calculation(see appendix) which shows that at $\ga = 1/2$ we indeed have the minimum with  $\om = \om_{\mbox{\tiny BFKL}}$.

  \subsection{Independence of the Pomeron spectrum from the gluon mass}
 
 In this section we wish to prove that there are no Pomeron states above the intercept of the massless BFKL equation. As we have seen in the variational approach, the best trial function that describes the BFKL Pomeron  takes the form
 \beq \label{MIPS1}
 \phi^0_{\mbox{\tiny trial}}\,\,=\,\,\frac{1}{\sqrt{\kappa \,\,+\,\,a^2}}
 \eeq
 It gives  $E_{\mbox{\tiny BFKL}}\,\,=\,\,-\,4\,\ln 2$ independently from $a$ (see \fig{var}-a). We wish to prove that
 \beq \label{MIPS2}
 E\,\,\ge \,\,E_{\mbox{\tiny BFKL}}\,\,=\,\, -\,4\,\ln 2
 \eeq
 Since the energy contribution of the contact term is positive, we neglect it  below.
 
 For the proof of~(\ref{MIPS2}) we re-write the Hamiltonian of \eq{H1} in the form
 \beq \label{MIPS3}
 {\cal H}\,\,=\,\,T\Lb p \Rb\,\,+\,\,V\Lb r \Rb\,\,=\,\,\Big\{ T\Lb p \Rb\,\,-\,\, T_0\Lb p \Rb  \Big\}\,\,+\,\,{\cal  H}_0
 \eeq
 where ${\cal H}_0$ is chosen from the condition 
 
 \beq \label{MIPS4}
 {\cal  H}_0  \phi^0_{\mbox{\tiny trial}}\,\,=\,\,\Big( T_0\Lb p \Rb\,\,+\,\,V\Lb r \Rb\Big)\phi^0_{\mbox{\tiny trial}}\,\,=\,\, E_{\mbox{\tiny BFKL}} \phi^0_{\mbox{\tiny trial}}
 \eeq
 
 If we verify that $\{ T(p)-T_0(p)\}\ge 0$ for all values of $p$, then inequality~(\ref{MIPS2}) is valid due to~(\ref{MIPS4}) because $\phi ^0_{trial}$ is positive for the ground state of $H_0$.
 
Neglecting the contact term, the kinetic energy $T_0(p)$ takes the form\footnote{The ordering in \eq{MIPS5} is essential since $\phi^0_{\mbox{\tiny trial}}\Lb p \Rb$ is an operator in coordinate space.}
\beq \label{MIPS5}
T_0\Lb p\Rb\,\,=\,\,E_{\mbox{\tiny BFKL}} \,\,-\,\,\frac{1}{\phi^0_{\mbox{\tiny trial}}\Lb p \Rb}V(r)\phi^0_{\mbox{\tiny trial}}\Lb r \Rb\,
\eeq
where
\beq \label{MIPS6}
\frac{1}{\phi^0_{\mbox{\tiny trial}}\Lb p \Rb}V(r)\phi^0_{\mbox{\tiny trial}}\Lb r \Rb\,=\, - \int \frac{d^2p'}{\pi}\frac{\sqrt{p^{ 2}+a^2}}{ (|\vec{p} - \vec{p}^{\,'}|^2+1)\sqrt{p^{\prime 2}+a^2}}
=-\int_0^1\frac{d \beta}{\sqrt{1-\beta}}\, \frac{\sqrt{p^{ 2}+a^2}}{\sqrt{\beta (1-\beta )p^{ 2}+a^2(1-\beta )+\beta }}\,\eeq
The last expression can be written in terms of the elliptic integral in the Weierstrass form or in the Jacobi form
after the following transformation
\beq \label{MIPS7}
\frac{1}{\phi^0_{\mbox{\tiny trial}}\Lb p \Rb}V(r)\phi^0_{\mbox{\tiny trial}}\Lb r \Rb\,\,=\,\,-2 \int_0^1 d z \frac{\sqrt{p^{ 2}+a^2}}{\sqrt{z^2(1-z^2) p^{ 2}+a^2\,z^2+1-z^2}}\,.
\eeq
 
 For \eq{MIPS7} we can find the asymptotic behaviour for large and  small $p$, viz.
 
 \bea 
 \frac{1}{\phi^0_{\mbox{\tiny trial}}\Lb p \Rb}V(r)\phi^0_{\mbox{\tiny trial}}\Lb r \Rb \,\,~&\xrightarrow{ p\,\,\gg\,\,1}&~~- \frac{2}{p}\,\ln p-4\ln 2+\frac{\ln p}{p^2}\,\left(-\frac{1}{2}+a\right)+{\cal O}(1/p^2);\label{LAP}\\
 &\xrightarrow{ p\,\,\to\,\,0}&~~  -\,\frac{2\sqrt{a^2}}{\sqrt{a^2\,-\,1}}\,
\ln (\sqrt{a^2\,-\,1}+\sqrt{a^2}) \label{SMP}
 \eea
 
 In terms of $\Big\{ T\Lb p \Rb\,\,-\,\, T_0\Lb p \Rb  \Big\}$ it means that
 \bea 
 \Big\{ T\Lb p \Rb\,\,-\,\, T_0\Lb p \Rb  \Big\} ~~&\xrightarrow{p \,\gg\,1}&~~\frac{\ln p}{p}\,\left(a-\frac{5}{2}\right); \label{LAPT}
 \\
 ~~&\xrightarrow{p \,\to\,0}&~~ \frac{1}{2}+4\ln 2 \,-\,2\sqrt{\frac{a^2}{a^2-1}}\,\ln (\sqrt{a^2-1}+\sqrt{a^2}).\label{SMPT}
 \eea
 As a result, it is plausible, that $T(p)-T_0(p)$ is positive for all $p$ providing that the parameter $a$
lies in the interval
\beq \label{MIPS8}
\frac{5}{2}\, <\,a^2\, <\,a^2_0\,,
\eeq
where $a_0$
is found from the equation 
\beq \label{MIPS9}
\frac{1}{2}\,+\,4\ln 2\, \,-2\sqrt{\frac{a^2_0}{a^2_0-1}}\,\ln (\sqrt{a^2_0-1}+\sqrt{a^2_0})=0\,.
\eeq
which gives $a^2_0 = 5.26$.

\begin{figure}[ht]
     \begin{center}
     \includegraphics[width=10cm]{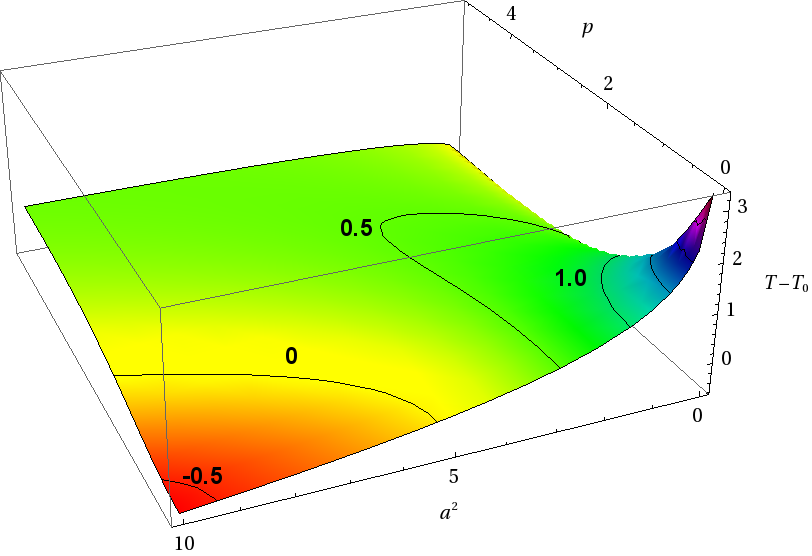} 
     \end{center}    
      \caption{ The dependence of  $ T\Lb p \Rb\,\,-\,\, T_0\Lb p \Rb$  on the values of parameter $a$ and $ p$. }
\label{VTD}
   \end{figure}

In \fig{VTD} we calculated the difference $ T\Lb p \Rb\,\,-\,\, T_0\Lb p \Rb$ using the integral of \eq{MIPS6} and/or \eq{MIPS7} without expansion of \eq{LAPT} and \eq{SMPT}. One can see that for $ 5 \,>\,a^2 \,>\,0$ at any values of $p$ this difference is positive.

The condition of the minimum of $|T(p)-T_0(p)|$ should be used in the variational approach for
fixing the unique wave function, because the minimum of energy is realized on many configurations.

\begin{figure}[ht]
     \begin{center}
     \includegraphics[width=7cm]{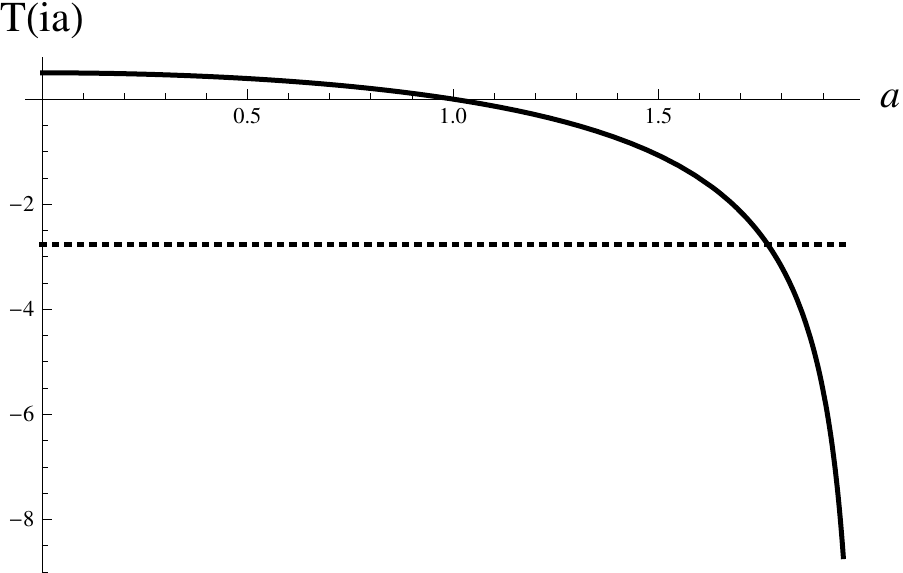} 
     \end{center}    
      \caption{ The dependence of  $ T\Lb i a \Rb $  on the values of parameter $a$ (solid line) and $E = -\chi(0) = - 4\ln 2$ (dotted line). }
\label{ta}
   \end{figure}

\fig{ta} shows that the condition of \eq{ABEQ}: $E = T\Lb p \Rb = T\Lb i a \Rb$,  is fulfilled for $a$ in the interval of \eq{MIPS8}(or \fig{VTD}). Thus, inequality~(\ref{MIPS2}) is proven.

  \subsection{ Relation between energy and wave function }
  
  In this section we  demonstrate that the value of energy  $E\Lb \beta \Rb$  is completely determined by the asymptotic
behavior of the wave function at large $ p$ for a more general trial function of the form
  
  \beq	 \label{REWA1}
 \phi_{\mbox{trial}}\Lb p \Rb  \,\,=\,\,\Lb p^2\,\,+\,\,a^2\Rb^{ - \h\,+\,i \beta}.
 \eeq 
  This proof complements the proof given in section 3.1, in which we used properties of the massless BFKL equation and argued that the spectrum of massless and massive BFKL kernels  should coincide at large $p$. The trial function~\eq{REWA1} is close to the wave functions which we will obtain numerically in section 3.6, so we find it instructive to  repeat the proof for these functions in a more transparent way.
  
  For the trial function of \eq{REWA1} \eq{MIPS8} takes the form
   \bea \label{REWA2}
   \frac{V(r)\phi_{\mbox{\tiny trial}}\Lb r \Rb}{\phi_{\mbox{\tiny trial}}\Lb p \Rb}\,&=&\\
&- & \int \frac{d^2p'}{\pi}\frac{\Lb p^{ 2}+a^2\Rb^{\h + i  \beta}}{ (|\vec{p} - \vec{p}^{\,'}|^2+1)\Lb p^{\prime 2}+a^2\Rb^{\h + i \beta}}
=-\int_0^1\frac{d x}{\sqrt{1-\beta}}\, \frac{\Lb p^{ 2}+a^2\Rb^{\h + i \beta}}{\Lb \beta (1- x )p^{ 2}+a^2(1- x )+ x \Rb^{\h + i \beta}}\nn
\eea  
We introduced Feynman parameter $x$ and integrated over $p'$ to obtain the last equation in~\eq{REWA2}.

For large $p$  the essential region of integration is $ a^2/p^2  \leq  x \leq 1$. We introduce an
intermediate parameter $\sigma$ with its value in the interval 
  $ a^2/p^2\,\ll\,\sigma\,\ll\,1$ and rewrite \eq{REWA2} in the form
\bea
&&\frac{V(r)\phi_{\mbox{\tiny trial}}\Lb r \Rb}{\phi_{\mbox{\tiny trial}}\Lb p \Rb}\,= \\
&& - \int^\sigma_0 \frac{d x}{x} \frac{1}{\Lb 1 + \frac{1}{x p^2}\Rb^{\h + i \beta}}\,\,-\,\,\int^1_\sigma \frac{d x}{x} \frac{1}{\Lb 1 - x\Rb^{\h + i \beta}}\,\,
 = \,\,- \int^\infty_{1 +  \frac{1}{\sigma p^2}}\frac{d z}{\Lb  z - 1\Rb}  \,\,\frac{1}{z^{\h + i \beta}}\,\,-\,\,\int^1_\sigma \frac{d x}{x} \frac{1}{\Lb 1 - x\Rb^{\h + i \beta}}\nn\\
 &&=\,\, - \int^{ 1 - \frac{1}{\sigma p^2}}_0 \frac{d t}{t (1 - t)}\,t^{\h + i \beta}\,\,-\,\,\int^{1 - \sigma}_0 \frac{d t}{1 - t} \,t^{-\h - i \beta}\,\,
 = \,\,- \ln p^2\,\,-\,\,\int^1_0 \,d t \frac{t^{-\h}\Lb t^{- i \beta} + t^{ i \beta}\Rb - 2}{(1 - t)}\nn
 \eea

Therefore,
\beq \label{REWA3}
E\Lb \beta \Rb\,\,=\,\,\psi\Lb \h + i \beta\Rb\,\,+\,\,\psi\Lb \h - i \beta\Rb\,\,-\,\,2\,\psi\Lb 1\Rb
\eeq
independently of the value of $a$. Moreover, the result for the energy $E\Lb \beta \Rb$  does not depend
on the form of wave function providing that it has the correct asymptotic behavior
at large $p$ . For example,  the wave function $\phi^{(approx)}_n\Lb \kappa\Rb$ of \eq{PARWF} that stems from our numerical estimates, can be written as the real part of the expression
\beq \label{REWA4}
\phi^{(approx)}_n\Lb \kappa\Rb\,\,=\,\,\frac{e^{i \varphi}}{\sqrt{\kappa + 4}}\Lb \frac{\sqrt{\kappa + 4} + \sqrt{\kappa}}{\sqrt{\kappa + 4} - \sqrt{\kappa}}\Rb^{- i \beta}
\eeq
The difference of energy for the wave functions of \eq{REWA1} and \eq{REWA4}  takes the form
\beq \label{REWA5}
\Delta E\Lb \beta \Rb\,\,=\,\,\int \frac{d ^2 p'}{\pi} \frac{ p^{2( \h + i \beta)}}{\Lb | \vec{p} - \vec{p}^{\,'}|^2 + 1\Rb\sqrt{p'^2 + 4}}\Lb \frac{1}{\Lb p'^2 + 4\Rb^{i \beta}}\,\,-\,\,\Lb \frac{\sqrt{p'^2 + 4} + \sqrt{p'^2}}{\sqrt{p'^2 + 4} - \sqrt{p'^2}}\Rb^{- i \beta}\Rb
\eeq
From the dimensional considerations  $\Delta E \Lb \beta\Rb$ falls down as $1/p^2$  at large $p$ and therefore, the energies  $E\Lb \beta \Rb$ for  wave function of \eq{REWA1} and \eq{REWA4} coincide.

\subsection{Numerical solution}
 
\subsubsection{Direct method}
\paragraph{General approach.}

 \eq{EQF} and \eq{EQBFKL} have the following structure
\beq \label{NS1}
\om\,\phi\Lb \kappa\Rb\,\,=\,\,\bas\,\int d \kappa' K\Lb \kappa, \kappa'\Rb \,\phi\Lb \kappa'\Rb
\eeq
Notice that we re-write \eq{EQF} and \eq{EQBFKL} in terms of $\om$ and  restore the coupling constant in front of the integral.
In the numerical calculation we replace the continuous variables $\kappa$ and $\kappa'$ by the discrete set of $\{\kappa_n\}$ and
 $\{\kappa'_n\}$ using the logarithmic grid
(in $\kappa=k^{2}/m^{2}$) with $N+1$ nodes, 
\bea\label{NS2}
\kappa_{n} & =&\kappa_{min}\exp\left(\frac{n}{N}\,\ln\left(\kappa_{max}/\kappa_{min}\right)\right),\quad n=0,...,N,
\eea
where the values of $\kappa_{min},\,\kappa_{max}$ were set to $\kappa_{min}=10^{-40},\kappa_{max}=10^{80}$,
and $N=1024$.

\begin{figure}
\centerline{\includegraphics[scale=0.8]{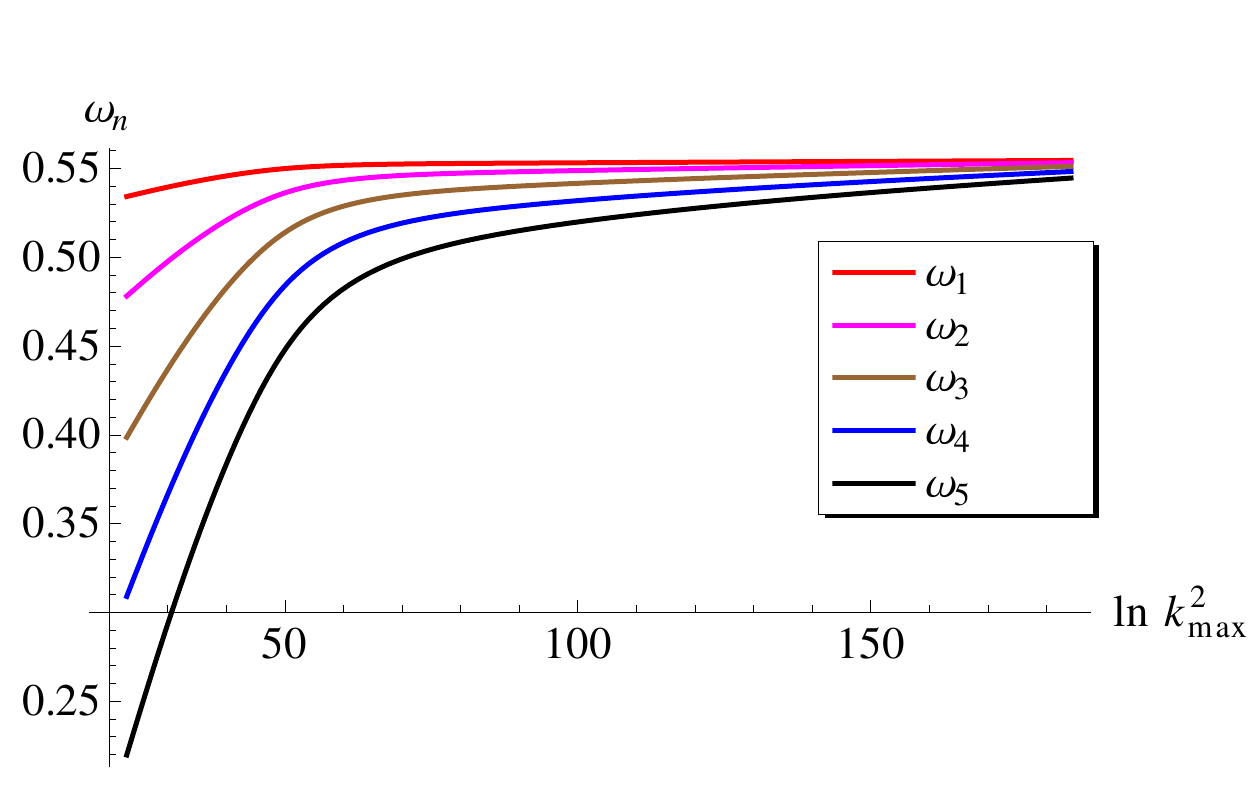}}
\caption{\label{directRoots-pMax}Dependence of the first 5 eigenvalues
on the maximal cutoff $\kappa_{max}=k_{max}^{2}$.}
\end{figure}

In the discrete  variables  \eq{NS1} takes the form
\beq \label{NS3}
\om \phi\Lb \kappa_n\Rb\,\,=\,\,\bas\sum^{N}_{m=0} K\Lb \kappa_n, \kappa'_m\Rb\,\kappa'_m\,\Lb\frac{1}{N}\,\ln\left(\kappa_{max}/\kappa_{min}\right)\Rb\,\phi\Lb\, \kappa'_m\Rb
\eeq
where $\kappa_n$ and $\kappa'_m$ are taken in the form of \eq{NS2}. Introducing the notations: $\phi\Lb \kappa_n\Rb \equiv \phi_n$ and $  K\Lb \kappa_n, \kappa'_m\Rb\kappa'_m\,\Lb\frac{1}{N}\,\ln\left(\kappa_{max}/\kappa_{min}\right)\Rb\,\equiv\,{\cal K}_{n m}$ we can re-write \eq{NS3} in the matrix form
\beq \label{NS4}
\om\,\phi_n\,\,=\,\,\bas \sum^{N}_{m=0} {\cal K}_{n m}\,\phi_m~~~~~~~~\mbox{or}~~~~~~~~~\omega\,\vec{\phi}\,\,=\,\,\bas \,{\cal \mathbf K}\,\vec{\phi}
\eeq
where vector $\vec{\phi}$ has $N+1$ components $\phi_n$ and ${\cal \mathbf K}$ is $(N+1) \times (N+1)$ matrix. To find the roots of the  characteristic polynomial $p\Lb \om\Rb$ of the matrix $\bas\,{\cal \mathbf K}\,-\,\om {\mathbf I}$, where ${\mathbf I} $ is the identity matrix, we need to solve the secular equation
\beq \label{NS5}
p\Lb \om\Rb\,\,=\,\,\mbox{det}\Lb \bas\, {\cal \mathbf K}\,-\,\om {\mathbf I}\Rb\,\,=\,\,0
\eeq
We use~\eq{NS4} and \eq{NS5} to find the eigenvalues and eigenfunctions both for massive~(\ref{EQF}) and massless~(\ref{EQBFKL}) BFKL equations, using the analytic solution~\eq{LKAP} to control the accuracy of our numerical calculations. Due to finite grid size, the spectrum is discrete, with a few positive
roots given in the Table~\ref{tab:dircetBFKL} and \fig{directRoots-pMax}.  Sensitivity to a number of points is quite mild, so discretization error should be small. As one can see from the \fig{directRoots-pMax}, when $\kappa_{max}$ grows up to infinity, the distance between the roots decreases rapidly, with the highest root asymptotically approaching the massless BFKL value $\om_{\mbox{\tiny BFKL}}\, =\,4\, \bas\,\ln2\,\,\,\approx\,\,0.56$ for $\bas = 0.2$, both for the massive and massless cases. It should be stressed that the relative difference between the highest eigenvalue in our calculation for the massless BFKL equation and the exact $\omega_{BFKL}$ is negligibly small (of the order of  $3\times10^{-5}$), which demonstrates a good accuracy of a chosen method.
We found that the eigenvalues of the massless BFKL equation can be written in a familiar form
\bea \label{EVN0}
\omega_n\Lb m=0 \Rb\,\,&=&\,\,\bas \Lb 2 \psi\Lb 1 \Rb \,-\,\psi\Lb \h - i \beta_n(m=0)\Rb \,-\,\psi\Lb \h + i \beta_n(m=0)\Rb\Rb\\
\mbox{with}&&~~~~~ \beta_n(m) \,=\,a \Lb m \Rb \,n,       ~~~~~\qquad ~~~~\,a\Lb m \Rb \,=\,2.9 / \ln \Lb \kappa_{\mbox{max}}/\Lb \kappa_{\mbox{min}}\,+\,m^2\Rb\Rb,\nn
\eea
and $\kappa_{\mbox{max}} $ and $\kappa_{\mbox{min}}$ 
are the upper and lower cutoffs introduced in~\eq{NS2}.

\begin{table}[h]
\global\long\def\arraystretch{1.5}
\begin{center}
\begin{tabular}{|c|c|c|c|c|c|c|}
\cline{1-3} \cline{5-7} 
Root \# & $\omega_{n}^{(BFKL)}$ & $\omega_{n}^{(mass)}$ &  & Root \# & $\omega_{n}^{(BFKL)}$ & $\omega_{n}^{(mass)}$\tabularnewline
\cline{1-3} \cline{5-7} 
1  & 0.5545  & 0.554  &  & 11  & 0.507  & 0.454 \tabularnewline
\cline{1-3} \cline{5-7} 
2  & 0.553  & 0.551  &  & 12  & 0.499  & 0.437 \tabularnewline
\cline{1-3} \cline{5-7} 
3  & 0.551  & 0.547  &  & 13  & 0.489  & 0.420 \tabularnewline
\cline{1-3} \cline{5-7} 
4  & 0.548  & 0.540  &  & 14  & 0.480  & 0.402 \tabularnewline
\cline{1-3} \cline{5-7} 
5  & 0.545  & 0.532  &  & 15  & 0.470  & 0.383 \tabularnewline
\cline{1-3} \cline{5-7} 
6  & 0.540  & 0.522  &  & 16  & 0.459  & 0.365 \tabularnewline
\cline{1-3} \cline{5-7} 
7  & 0.535  & 0.511  &  & 17  & 0.448  & 0.346 \tabularnewline
\cline{1-3} \cline{5-7} 
8  & 0.529  & 0.498  &  & 18  & 0.437  & 0.327 \tabularnewline
\cline{1-3} \cline{5-7} 
9  & 0.522  & 0.485  &  & 19  & 0.426  & 0.308 \tabularnewline
\cline{1-3} \cline{5-7} 
10  & 0.515  & 0.470  &  & 20  & 0.414  & 0.289 \tabularnewline
\cline{1-3} \cline{5-7} 
\end{tabular}
\end{center}
\caption{\label{tab:dircetBFKL}The first twenty roots of the original (massless)  BFKL equation
(column $\omega_{n}^{(BFKL)}$) and BFKL with mass (column $\omega_{n}^{(mass)}$)
found with the chosen method. Note that for the first root for the massless
BFKL, we get $\omega_{0}^{BFKL}\,\approx\,0.554504$, whereas the true
value is $4\bar{\alpha}\ln2\,\approx\, 0.554518$, i.e. the relative difference
is of order $3\times10^{-5}$.}
\end{table}

\begin{figure}
\centerline{\includegraphics[scale=0.8]{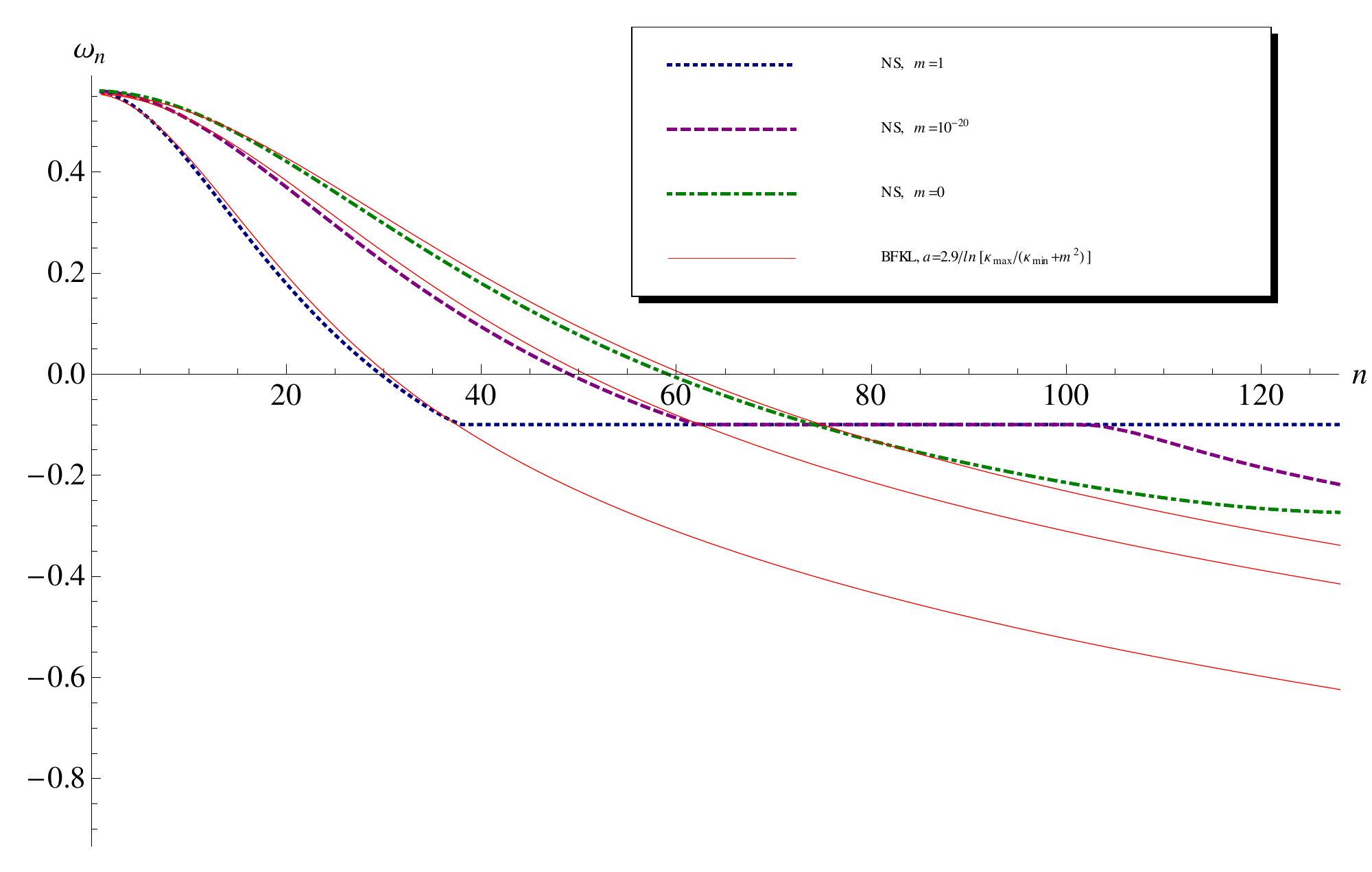}}
\caption{\label{omn} Dependence of the  eigenvalues on the number of zeros of the eigenfunction. The 
dashed lines describe the numerical solution for $\omega_n$ for the massive BFKL equation with mass $m$. $m=0$ corresponds to massless BFKL. The orange solid curves show the values of $\omega_n$
 calculated using \protect\eq{EVN0}
.}
\end{figure}

In \fig{omn} one can see how the simple formula of \eq{EVN0} describes the calculated spectrum (see solid and dashed curves for $m=0$).
For massive BFKL  situation is different. A simple parametrization~\eq{EVN0} with nonzero $m$  may be used with a good precision only for $\omega\geq\omega_0$. The point $\omega\approx \omega_0$ is special and will be discussed in more detail below. For very large $n$ the values of the intercepts become smaller than $\omega_0$ and agree with \eq{EVN0}, but the $n$-dependence of $\beta_n$ is no longer linear and will be discussed in the following section.

\paragraph{Eigenfunctions and Green's function.}

~

\begin{figure}
\begin{tabular}{ c c c}
\includegraphics[scale=0.5]{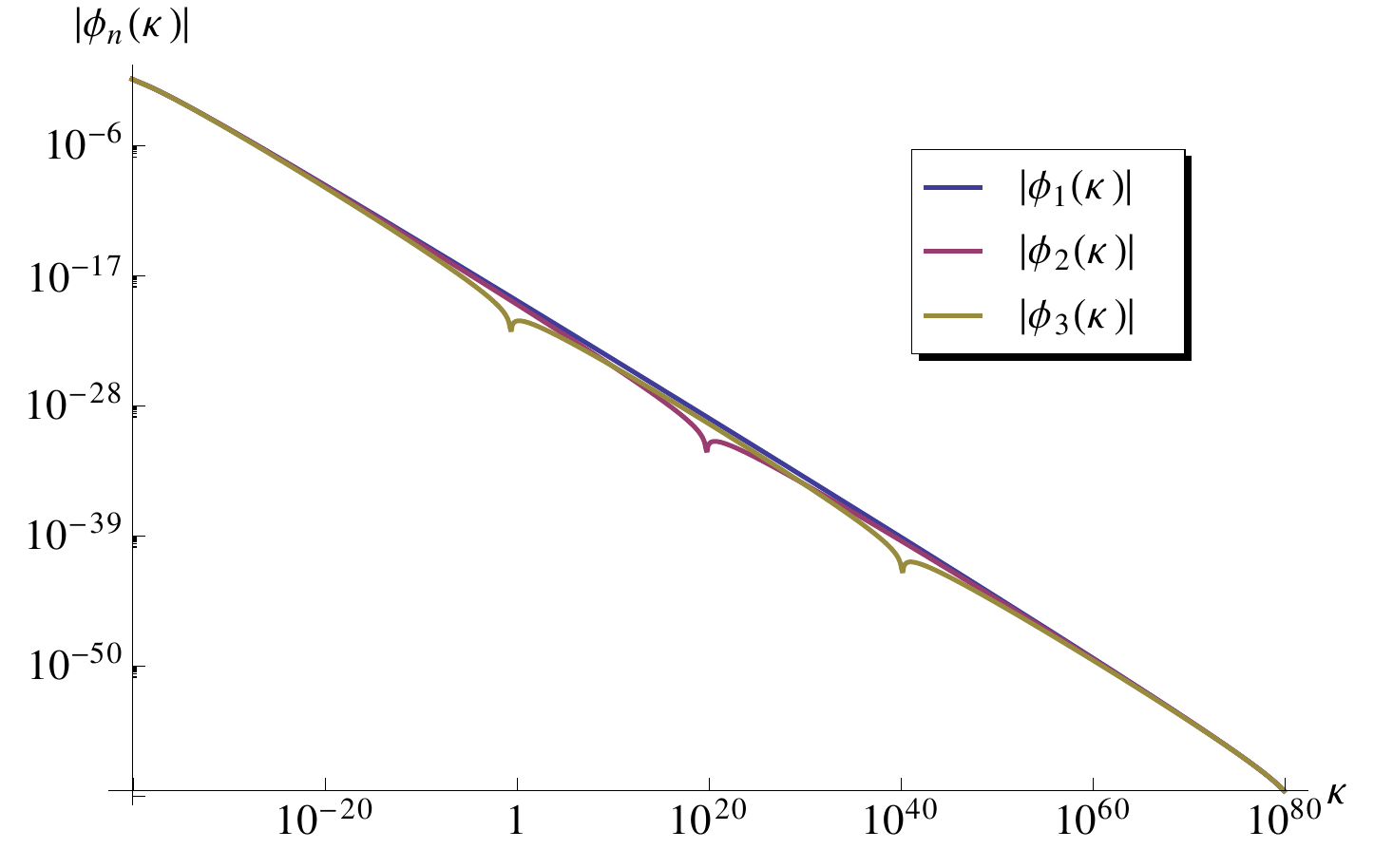}&~~~~~~~~~&\includegraphics[scale=0.37]{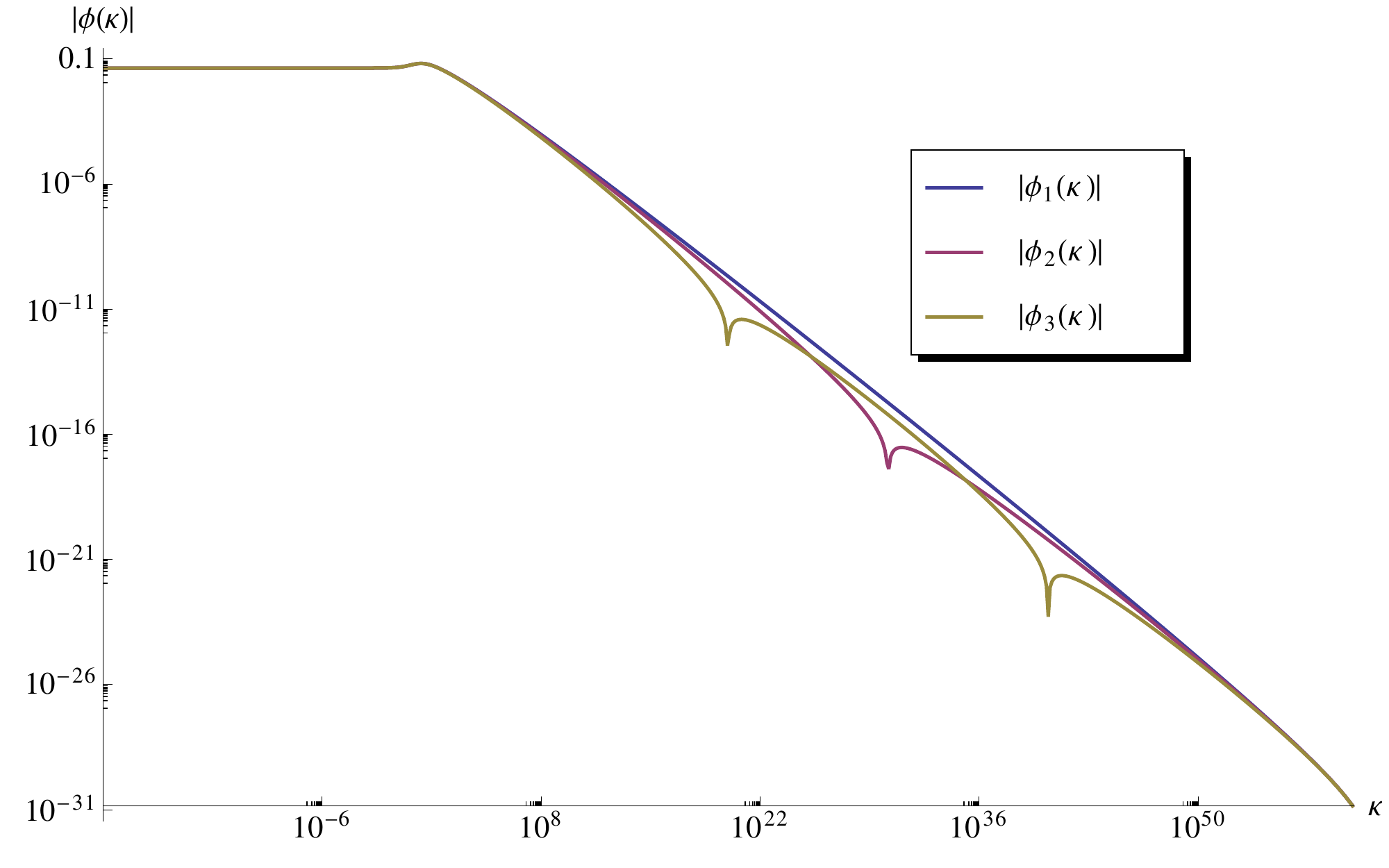}\\
\fig{directWFs}-a &  & \fig{directWFs}-b\\
\end{tabular}
\caption{\label{directWFs} Absolute values of the first three eigenfunctions
$\left|\psi_{n}\left(k\right)\right|$ corresponding to massless BFKL(\protect\fig{directWFs}-a)  and BFKL
with mass(\protect\fig{directWFs}-b). In the figure $k^2\,=\,\kappa$.}
\end{figure}

\begin{boldmath}
{\bf Eigenfunctions with}  $\omega \,\,\geq\,\,\omega_0$
\end{boldmath}
 first three (unnormalized) eigenfunctions corresponding to massless and massive BFKL
are shown in \fig{directWFs}. As we can see, for large $\kappa$ solutions of these equations coincide, however for $\kappa\,\lesssim1\,$ they are different:
the massless solution grows roughly as power of momenta, $\kappa^{-\gamma}$,
whereas the solution in the massive case is regular and reaches a constant. One can see that $\psi_2\Lb \kappa\Rb $ has one zero while
$\psi_3\Lb \kappa\Rb $ has two zeroes.
This behaviour of the wave functions has been expected from the 
general analysis of the solution ( see subsection 1 of this section).

With a good precision the eigenfunctions with the eigenvalues larger than $\omega_0$ can be parameterized as
\beq \label{PARWF}
\phi^{(approx)}_n\Lb \kappa\Rb\,\,=\,\,\frac{\alpha\Lb n \Rb}{\sqrt{\kappa + 4}}\sin\Big(\beta_n\Lb  m = 1\Rb\,Ln\Lb \kappa\Rb + \varphi_n\Big)~~~~~~\mbox{with}~~~Ln\Lb \kappa\Rb\,=\,\ln\Big(\frac{\sqrt{\kappa\,+\,4}\,\, +\,\, \sqrt{\kappa}}{\sqrt{\kappa\,+\,4} \,\,-\,\, \sqrt{\kappa}}\Big)
\eeq
The form of the parameterization in \eq{PARWF} is inspired by the expression for the gluon trajectory $\omega(\kappa)$. 

For $\kappa \ll 1$ and for $\kappa \gg 1$ the function~(\ref{PARWF}) has an asymptotic form
\bea \label{EFP}
\phi^{(approx)}_n\Lb \kappa\Rb\,\,=\,\,\left\{\begin{array}{l}\,\,\,\h\alpha\Lb n\Rb \sin\Lb \varphi_n\Rb\,\,\,\,\,\mbox{for}\,\,\,\kappa\,\ll \,1\,,\\ \\
\,\,\,\frac{\alpha\Lb n\Rb }{\sqrt{\kappa}}\sin\Lb \beta_n\Lb m = 1\Rb \ln \kappa \,+\,\varphi_n \Rb\,\,\,\,\,\mbox{for}\,\,\,\kappa\,\gg\,1\,. \end{array}
\right.
\eea

\begin{figure}
\begin{tabular}{c c c}
\includegraphics[scale=0.5]{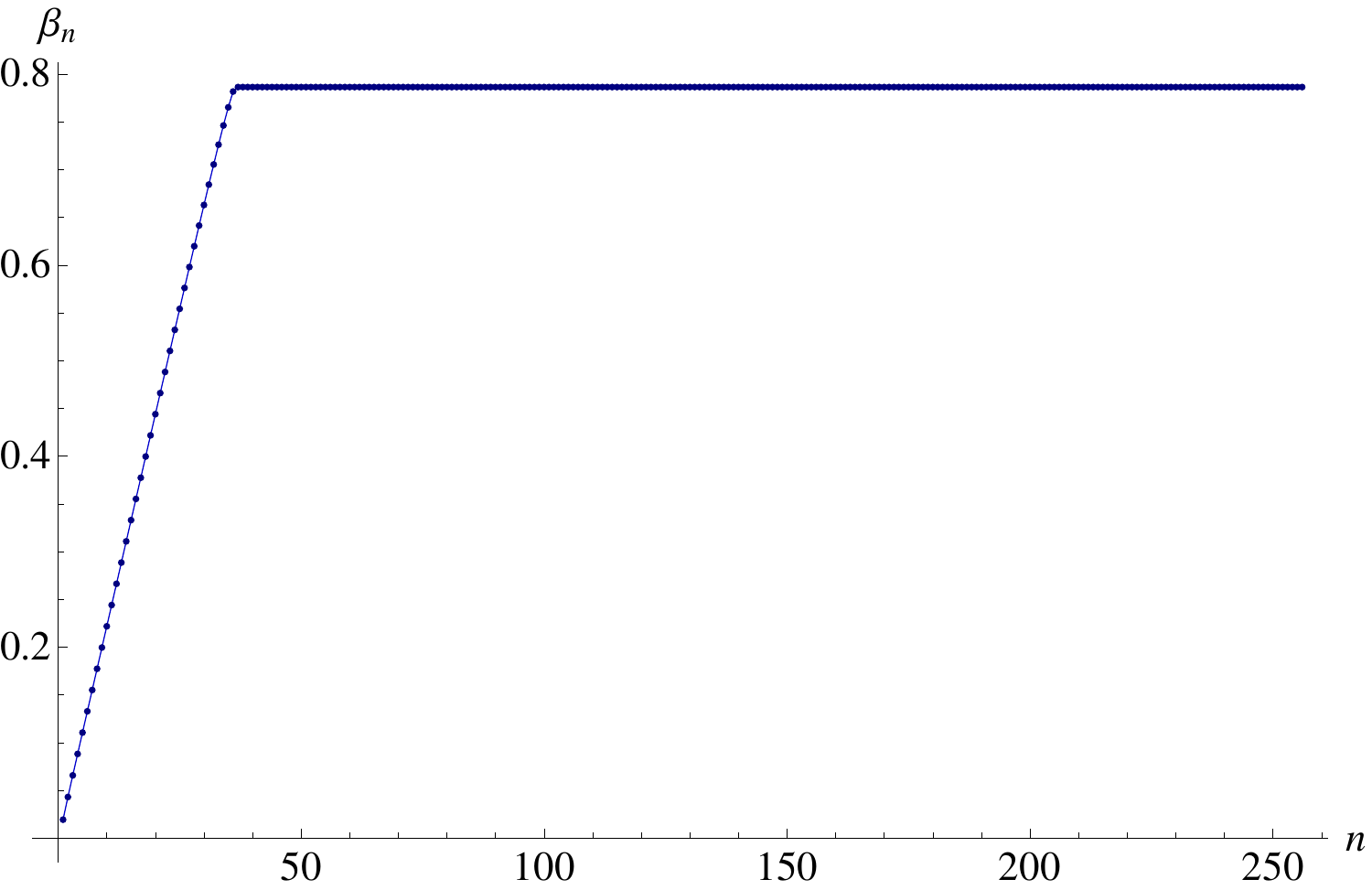} & ~~~~~&
\includegraphics[scale=0.4]{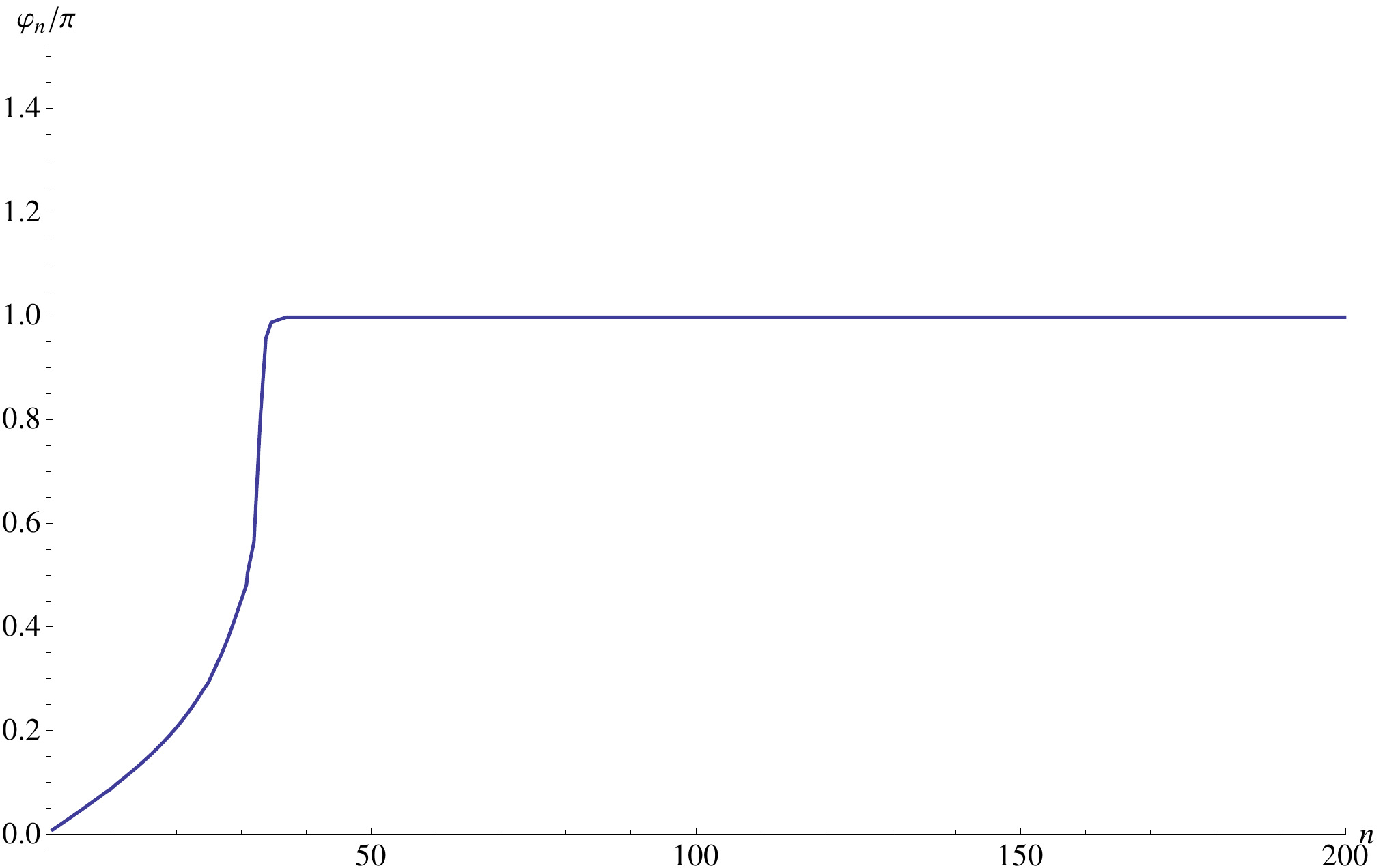}\\
\end{tabular}
\caption{\label{efp} Left: Dependence of the eigenfuncton parameter $\beta_n$ on eigenfunction number $n$. Right: Dependence of the eigenfuncton parameter $\varphi_n$ on eigenfunction number $n$. In a chosen lattice $\omega=\omega_0\equiv -\h \bas$ corresponds to $n=37$.}
\end{figure}

\begin{figure}
\center
\includegraphics[scale=0.45]{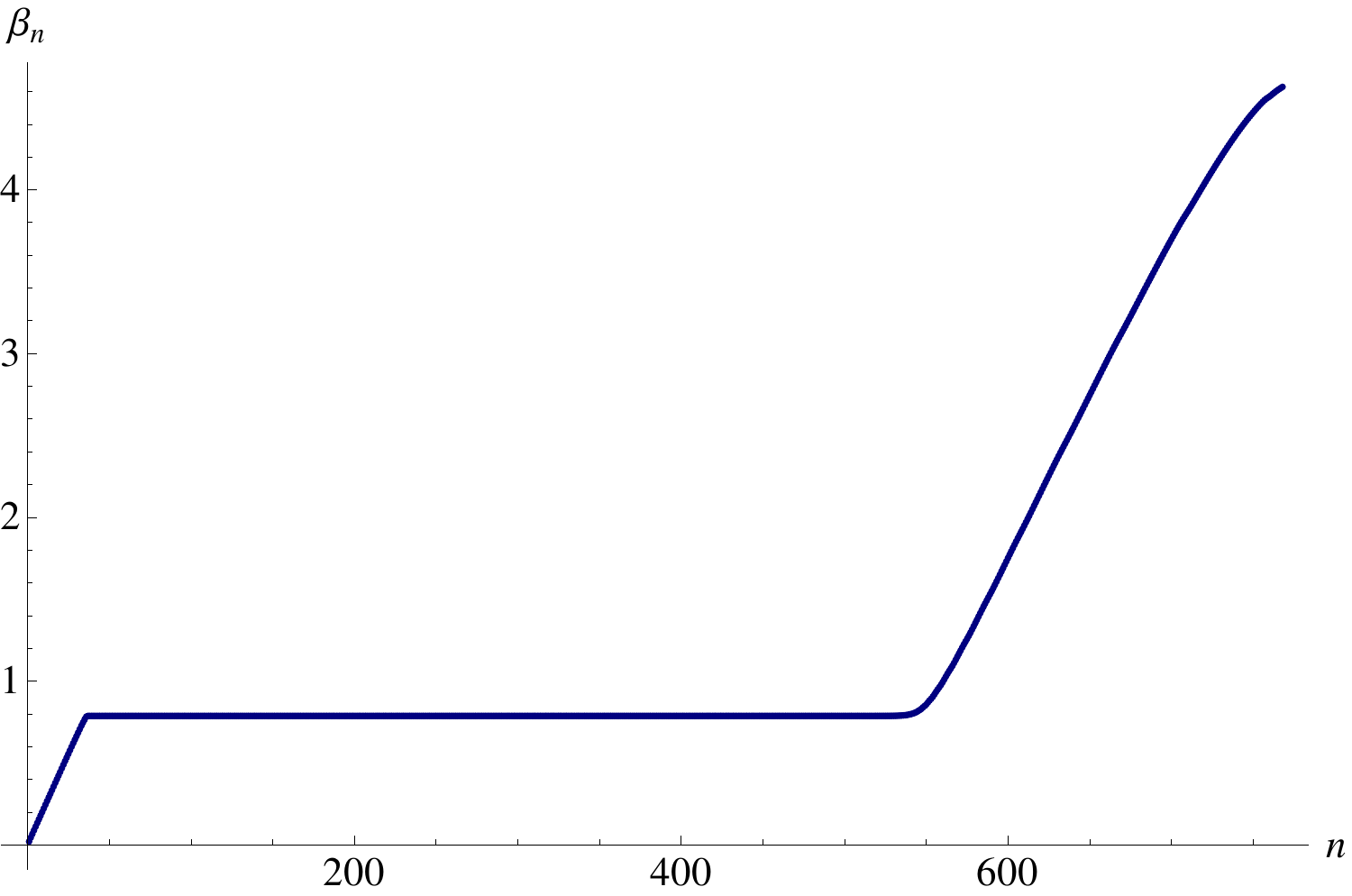}\\
\caption{ Large-$n$ dependence of the eigenfunction parameter $\beta$ on the eigenfunction number $n$\label{efpBeta}}
\end{figure}

Since in the large-$\kappa$ regime the massive BFKL coincides with massless BFKL, for which the second line of~\eq{EFP} is an exact solution, the parameters $\beta_n$ and $\varphi_n$ are defined for all possible values of $n$. For the case $m=1$ the dependence of $\beta_n$ and $\varphi_n$ on the number $n$  is shown in the~\fig{efp}. 
We can see that in the small-$n$ region both $\beta_n$ and $\varphi_n$ are linear functions of $n$,  
$\beta_n(m)\,=\,a (m)  n $ and $\varphi_n\,=\,a_\varphi(m) n$, where $a(m)$ is given by~\eq{EVN0}, and 
\bea
& a_\varphi(m) & \approx\frac{8.577\, }{\ln\left(\kappa_{max}/\kappa_{min}\right)}=b_\varphi \beta_n(m),\label{PhiP}\label{APHB1}\\
& b_\varphi & \approx 1.865,
\eea
so in this regime we may rewrite~\eq{PARWF} in a form
\bea
&\phi^{(approx)}\Lb \kappa,\beta\Rb\,\,=\,\,\frac{\alpha\Lb \beta \Rb}{\sqrt{\kappa + 4}}\sin\Big(\beta\,Ln\Lb \kappa\Rb + b_\varphi\,\beta \Big)&\label{APHB2}
\eea
which does not depend on lattice parameters. However, a linear approximation for $n$-dependence of $\phi_n$ is valid only for very small $n$. In the vicinity of the point $\omega=\omega_0$ both parameters freeze, and we'll discuss this regime in more detail in the next section. For very large $n$, the intercept $\omega$ goes below $\omega_0$  and the parameters $\beta$, $\varphi$ resume their dependence on $n$ (see e.g.~\fig{efpBeta}), however in this regime the oscillation period becomes comparable with period of the lattice, so extracted parameters are not very reliable. The normalization factor $\alpha(n)$ can be found from the normalization condition of~\eq{ORTCOM1} and is irrelevant for purposes of this paper since we are solving the linear equation.


\begin{figure}[ht]
\includegraphics[scale=0.6]{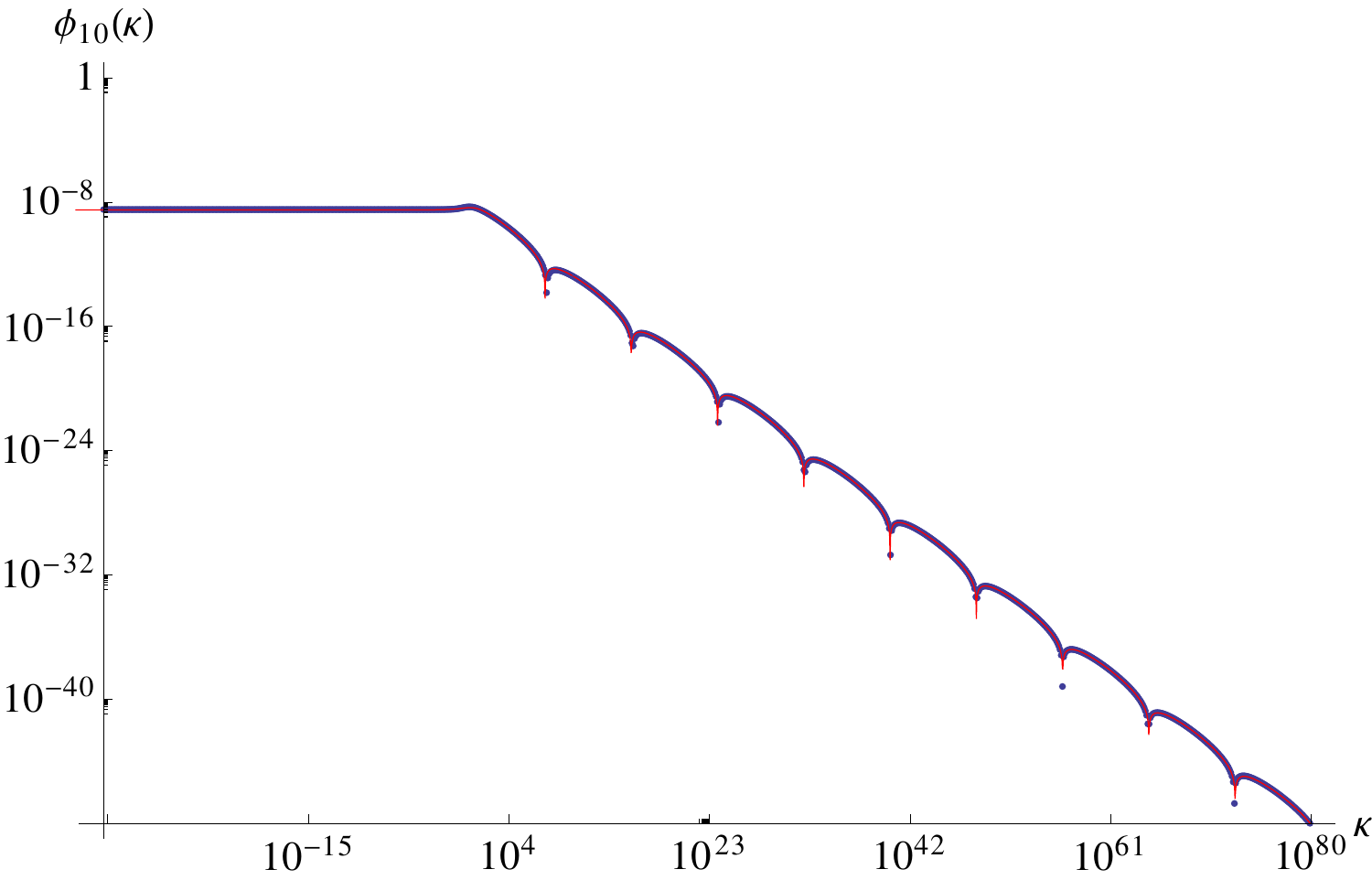}
\includegraphics[scale=0.68]{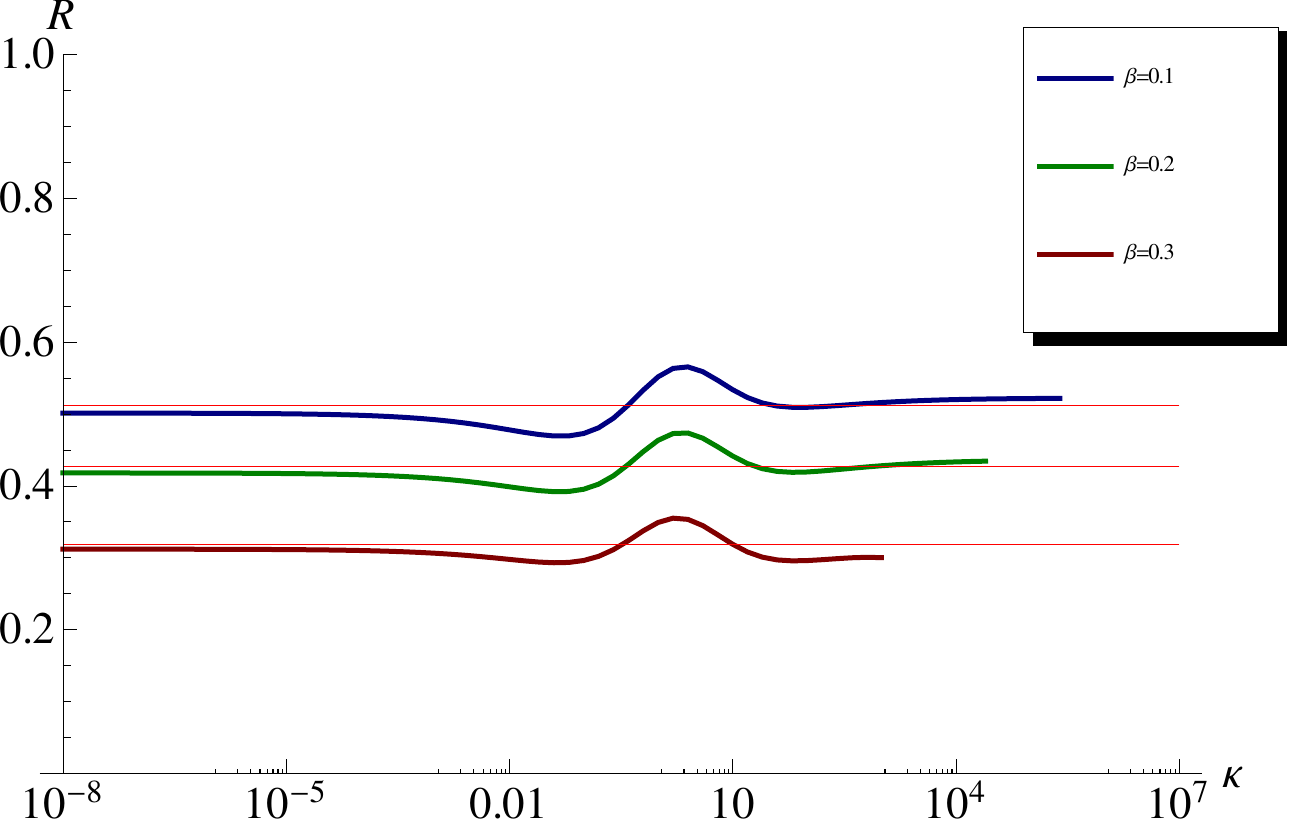}
\caption{\label{check} Left: Comparison of the approximate parametrization~(\protect\ref{PARWF}) (in red) with numerical result (blue) for $n=10$. Right: Check of accuracy of $\phi^{(approx)}\Lb \kappa,\beta\Rb$. $\omega^{approx} $  given by \protect\eq{CHECK} (see wavy lines) and $\omega^{exact}$   (orange straight lines) at different values of $\beta$. }
\end{figure}

In order to demonstrate the quality of the fit~(\ref{PARWF}), in the left pane of the~\fig{check} we directly compare the numerical eigefunction and parametrization~(\ref{PARWF}).  In the right pane of the~\fig{check} we plot the ratio
\beq \label{CHECK}
\omega^{approx}\,\,=\,\,\frac{\mbox{r.h.s.}\Lb  \phi^{(approx)}\Lb \kappa,\beta\Rb\Rb}{\phi^{(approx)}\Lb \kappa,\beta\Rb}
\eeq
which demonstrates that the deviations of the fit from numerical solution are the largest in the region $\kappa\sim 1$, however even there don't exceed 10\%.

~

\begin{boldmath}
{\bf  Eigenfunctions  in the vicinity of  $\omega \,\,=\,\omega_0$}
\end{boldmath}

As was discussed in previous sections, the point $\omega=\omega_0$ is special. We would like to investigate the behaviour near this point both analytically and numerically.
The equation of motion~\eq{EQF} for $\omega\approx \omega_0$, or $E=\h$ in the small-$\kappa$ regime has a form
  \beq \label{EQVE0}
   \Big( E - \h - \frac{5}{12}\,\kappa\Big)\phi\Lb \kappa \Rb\,\,=\,\,- \frac{ N^2_c - 1}{2 N^2_c}\int\frac{d \kappa'}{k' + 1}\,\phi\Lb \kappa'\Rb +\mathcal{O}\left(\kappa^2\right)
   \eeq
 Introducing a new notation $\epsilon = \frac{12}{5}\Lb E - \h\Rb$ one can see that function $\phi\Lb \kappa\Rb$ should have a pole at $\kappa=\epsilon$,
 \beq \label{PHI0}
 \phi\Lb  \kappa  \Rb|_{\kappa \to \epsilon \ll  1}\,\,=\,\,\frac{\rm const}{\epsilon\,-\,\kappa}.
 \eeq  
\begin{figure}[ht]
\centerline{\includegraphics[scale=0.4]{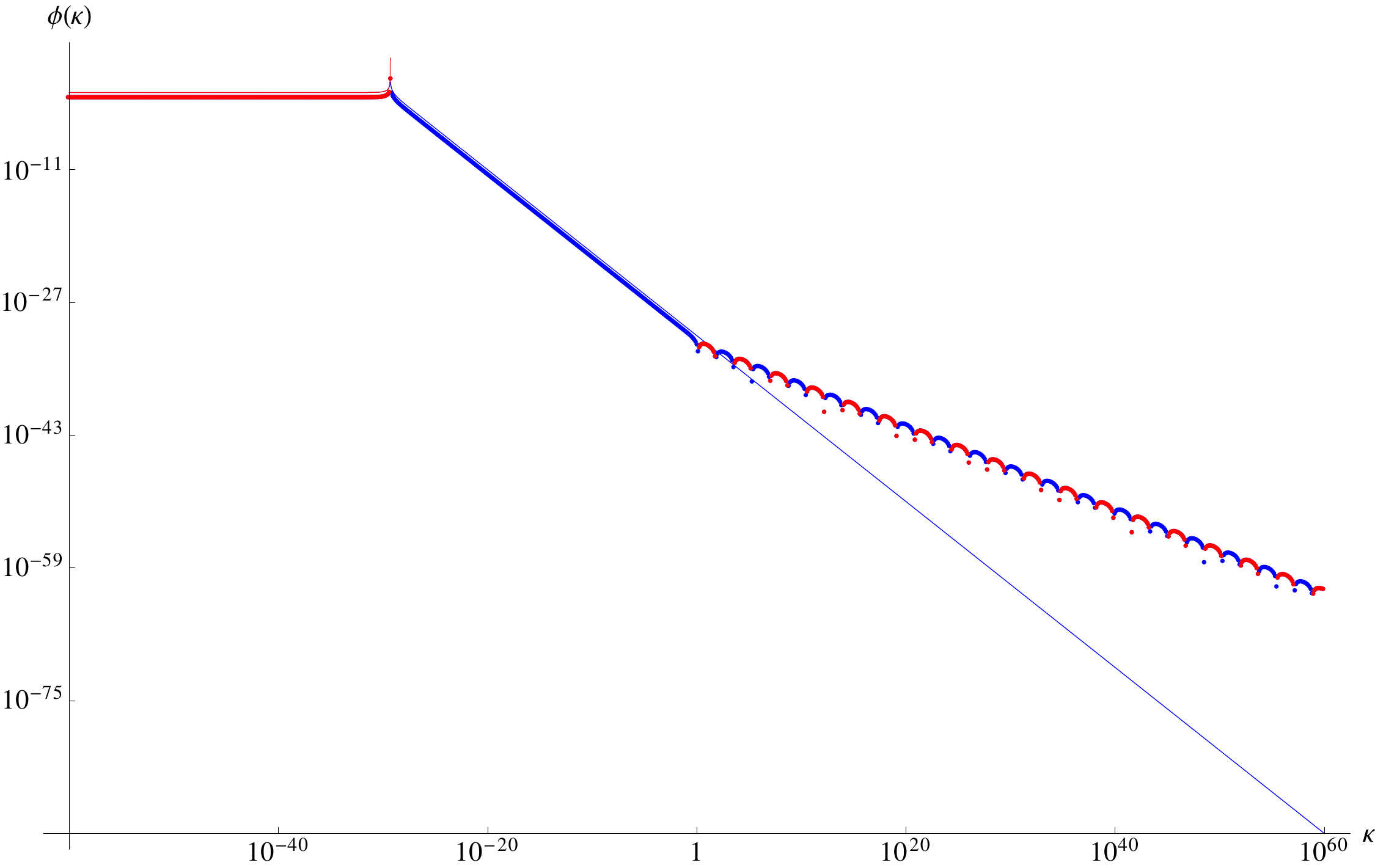}}
\caption{\label{pole} The eigenfunction $ | \phi (\omega_0, \kappa)|$  with $\omega = \omega_0\,=\, - \h \bas$.  Red curve corresponds to the positive values of  $ \phi (\omega_0, \kappa)$  while blue describes the negative $ \phi (\omega_0, \kappa)$. The approximate function $ | \phi^{approx} (\omega_0, \kappa)|=|1/\Lb \kappa\,-\,\kappa_0\Rb |$ with $ \kappa_0 = 10^{-30}$, is shown by the thin line.  We multiply the  fit result ($ \phi^{approx}$) by some constant to see the difference (otherwise they just coincide).
}
\end{figure}

A numerical calculation confirms our expectation. As one can see from~\fig{pole}, the wave function indeed has a pole at $\kappa\ll 1$. 
Position of the pole is arbitrary and may coincide with any node at $\kappa\ll 1$. Due to large number of nodes with $\kappa_n\ll 1$, the spectrum~\fig{omn} looks multiply degenerate at the point $\omega=\omega_0$. In order to demonstrate that this is not the case, we recalculated the eigenvalues in the lattice which has a linear step in the region $\kappa\leq 1$ and a logarithmic step for $\kappa\geq 1$, 
 \bea \label{LL}
\kappa_n=\,\,\left\{\begin{array}{l}\,\,\,\,\frac{n}{N_<}~~  \mbox{with}\,\,n = 0, \dots, N_<~~~N_<=200 ~~~\mbox{for}~~\kappa\,<\,1;\\ \\
\,\,\,\,\exp\Lb \frac{n}{N_>}\ln \kappa_{max}\Rb ~~\mbox{with}\,\,n = 0, \dots, N_>\,\,\,,\,
N_>=1024, \kappa_{max} = 10^{60},~~~\mbox{for}\,\,\,\kappa\,>\,1\,. \end{array}\right.
\eea

\begin{figure}[ht]
\centerline{\includegraphics[scale=0.65]{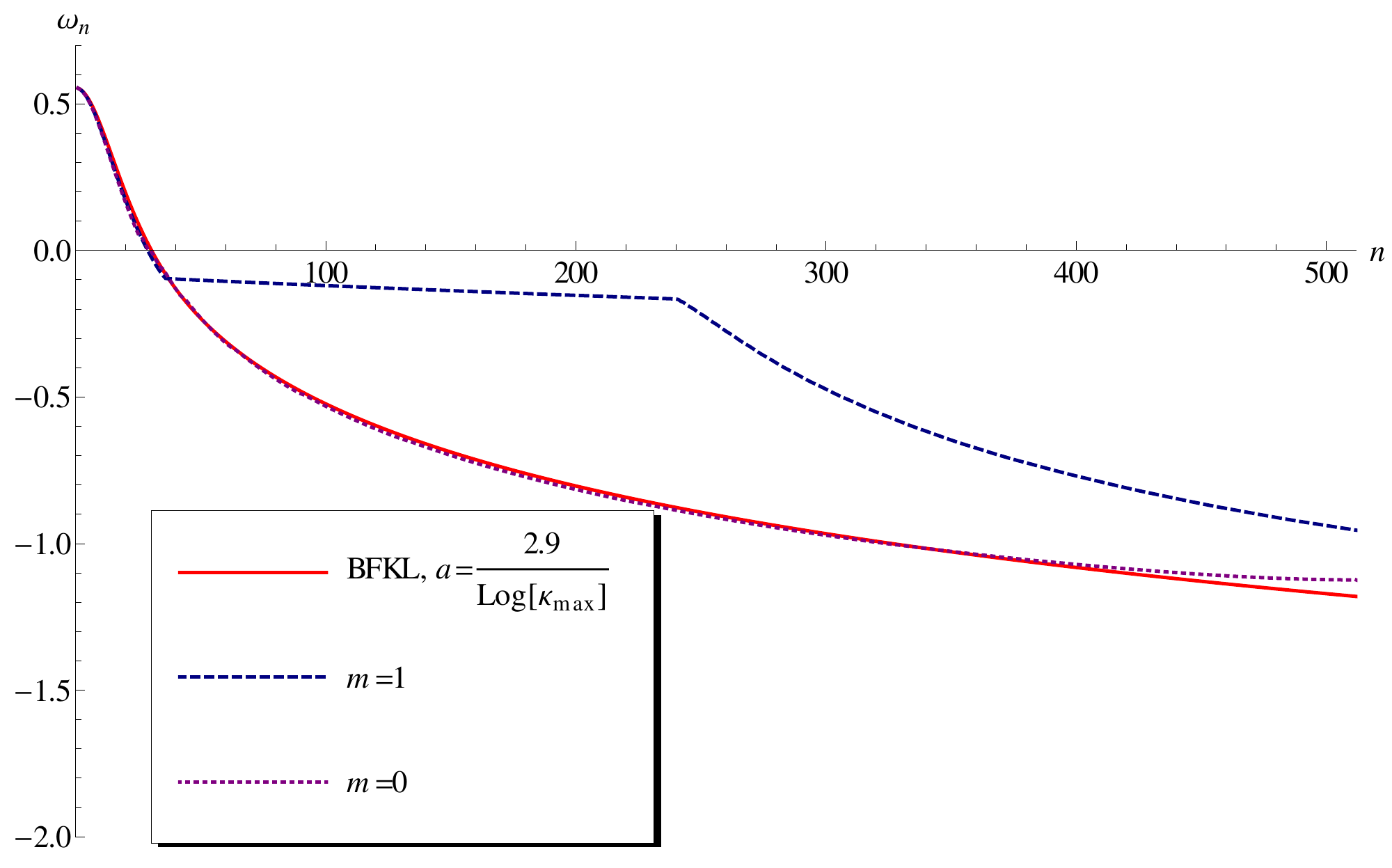}}
\caption{\label{ev} The  eigenvalues of the massless and massive BFKL equation in the linear-logarithmic discretization (see 
\protect \eq{LL}).The 
dashed lines describe the numerical solution for $\omega_n$ for the massive BFKL equation with mass $m$. $m=0$ corresponds to massless BFKL. The orange solid curves show the values of $\omega_n$
 calculated using \protect\eq{EVN0}
}
\end{figure}

From the~\fig{ev} we may see that the spectrum in this case is no longer degenerate. This happens because the typical node values $\kappa_n\sim 10^{-2}...10^{-1}$ are much larger than with logarithmic grid and a deviation $\omega-\omega_0\sim \kappa_n$ is also larger.

%

\begin{figure}[ht]
\begin{tabular}{ c c  c}
\includegraphics[scale=0.5]{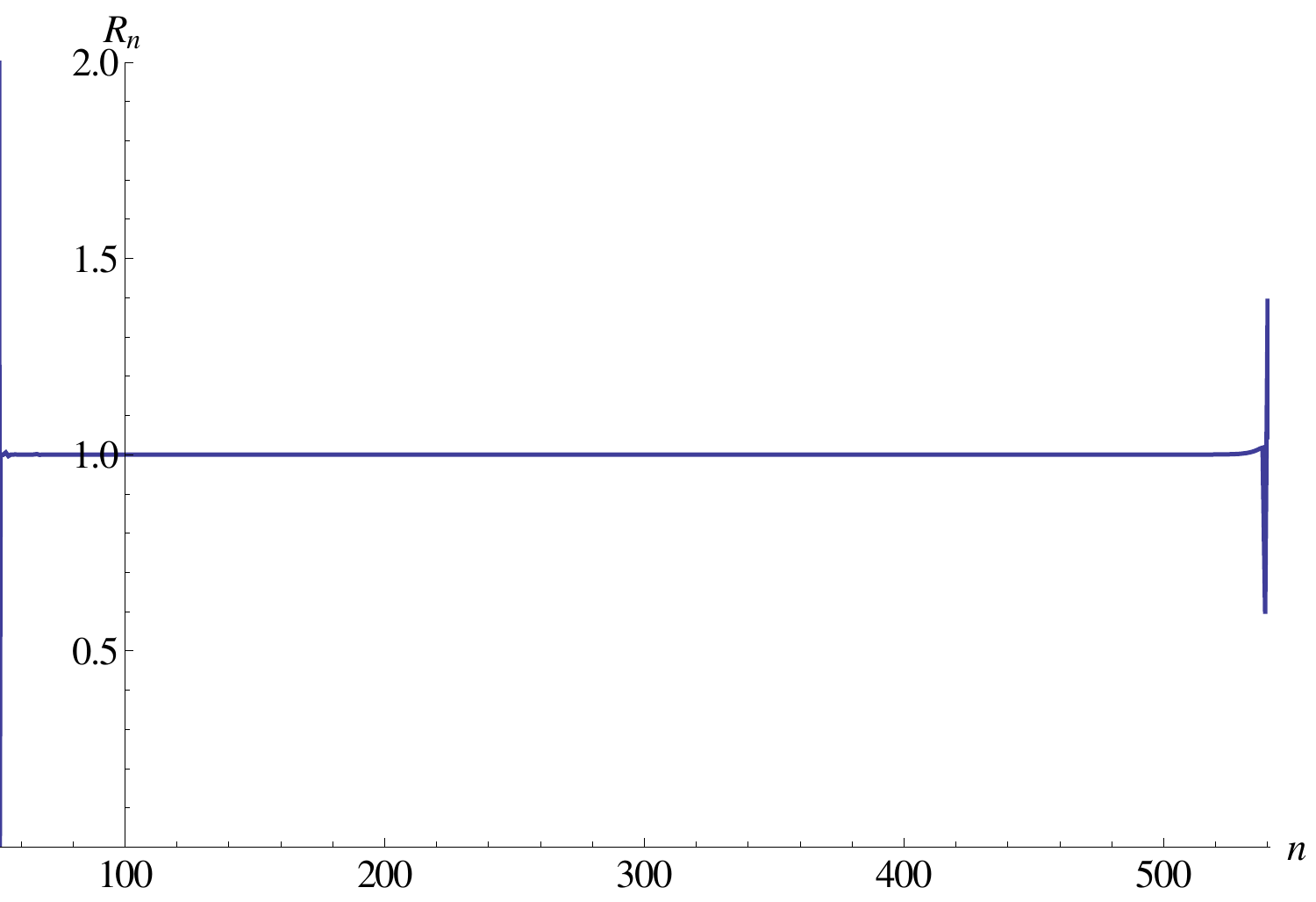}&~~~~~~~~~&\includegraphics[scale=0.5]{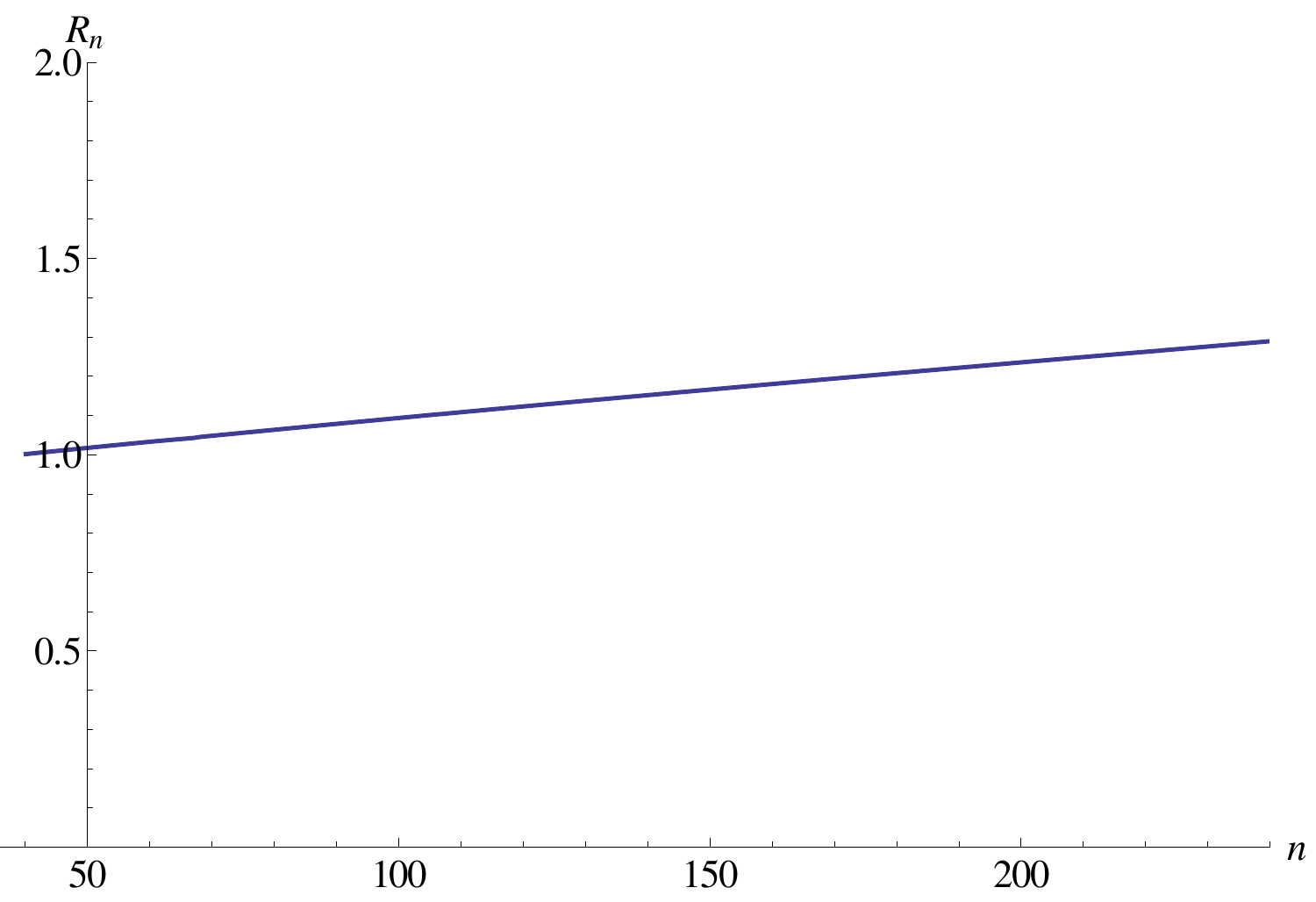}\\
\fig{coin}-a & &\fig{coin}-b\\
\end{tabular}
\caption{\label{coin} The ratio $R = \bas\Lb 5/12\Rb\kappa_0(n)/\Lb \om(n) - \om_0\Rb$ versus $n$. \protect\fig{coin}-a shows this ratio in logarithmic discretization (see  \protect\eq{NS2}, while in \protect\fig{coin}-b  the ratio is plotted in linear-logarithmic discretization (see \protect\eq{LL} for the description) }
\end{figure}

In the~\fig{coin} we demonstrate that the deviation $\omega(n)-\omega_0$ is proportional to the pole position $\kappa_0(n)$ and in agreement with~\eq{EQVE0} the ratio
$$R= \frac{5\bas}{12}\frac{\kappa_0(n)}{\omega_0(n) - \om_0}$$  is close to 1 for $\omega_0(n) \approx \omega_0$. In the left pane, we have shown results with logarithmic grid, and in the right pane  with grid~\eq{LL}. In the latter case, while results are close to one, there are some deviations due to $\mathcal{O}(\kappa^2)$ terms omitted in~\eq{EQVE0}.

For $\kappa > 1$, all the wave function in the vicinity $\omega\approx \omega_0$ have a form given by \eq{EFP} but with fixed $\beta_n = \beta^0 = 0.786$ found from  $\omega(\beta^0, m=1) = - \h \bas$, where $\omega(\beta^0, m=1)$ is given in~\eq{EVN0}.

 In summary, the wave functions with $\omega\approx \omega_0$ may be parametrized as
\bea \label{EFPB0}
\phi^{(approx)}_n\Lb \kappa\Rb\,\,=\,\,\left\{\begin{array}{l}\,\,\,\alpha\Lb n\Rb\,\sin \varphi_n \Lb1 - \kappa_0(n)\Rb /(\kappa - \kappa_0(n))\,\,\,\,\,\mbox{for}\,\,\,\kappa\,\leq \,1\,;\\ \\
\,\,\,\frac{\alpha\Lb n\Rb }{\sqrt{\kappa}}\sin\Lb \beta^0\, \ln \kappa \,+\,\varphi_n \Rb\,\,\,\,\,\mbox{for}\,\,\,\kappa\,>\,1\,; \end{array}
\right.
\eea

It is instructive to notice   that \eq{EFPB0} corresponds to the energy spectrum which  almost does not depend on $n$
for large range of $n$  independently of the type of discretization. This fact reflects in our calculation procedure the difference of the continuous spectrum  between $\omega\, >\,\omega_0$ and $\omega\, <\,\omega_0$. The former is discreet with the cut at large $\kappa$ while the latter remains continuous with this cut.


 \begin{boldmath}
{\bf  Eigenfunctions with $\omega \,\,<\,\,\omega_0$}
\end{boldmath}

~

For large $n $ ( $ n > 550$ see \fig{omn} and \fig{ev})  $\omega_n$ become smaller than $\omega_0$. In this kinematic region the eigenfunction can be described by general formulae of \eq{PARWF}  with $\beta$ that increases linearly with $n$ (see \fig{efpBeta}) but we need to add to this eigenfunction the term  $\propto 1/(\kappa - \kappa_0(n))$ with $\kappa_0(n) \,>\,1$.  However, the difference  $\Delta \kappa (n) = \kappa_0(n+1) - \kappa_0(n)$ turns out to be larger than
$\Delta E\Lb \beta(n)\Rb = E\Lb \beta(n + 1)\Rb - E\Lb \beta(n)\Rb$ of \eq{REWA3} for our discretization procedure. The appearance of  $1/(\kappa - \kappa_0(n))$ in the eigenfunction is the consequence of the fact that the spectrum remains continuous with the cut at large $\kappa$.

Note that evaluations in this region should be taken with due care because of possible interplay of oscillation period with period of the grid. The maximal value of $\beta$ which may be extracted with this method is controlled by the grid step and is given by $\beta_{max} =N/log(\kappa_{max}/\kappa_{min})~=\, 3.7$ (see \eq{NS2} for values of $N$, $\kappa_{max}$ and $\kappa_{min}$).

~

{\bf Green's function}

~
We  can calculate the Green function of the massive BFKL Pomeron using \eq{EFP}.  Indeed, the Green function takes the general form
\beq \label{GRFU}
G\Lb Y, \kappa_{fin} | 0, \kappa_{in}\Rb \,\,=\,\,\sum_{n=0}^{\infty} \phi_n\Lb \kappa_{fin}\Rb\, \phi_n\Lb \kappa_{in}\Rb\,e^{\omega\Lb n \Rb Y}
\eeq

where functions $\phi_n$ should be normalized according to   \eq{ORTCOM1}\footnote{In our numerical solution we have a discrete spectrum in the  restricted region of $\kappa$( from $\kappa_{min}$ to $ \kappa_{max}$). Therefore, we need to normalize not to $\delta$-function as in \eq{ORTCOM1} but to Kronecker's delta.}
In the diffusion approximation we can expand the eigenvalues of \eq{EVN0} at small $n$ replacing \eq{EVN0} by the simple expression
\beq  \label{EVND}
\omega\Lb n\Rb\,\,=\,\, \omega_{\mbox{\tiny BFKL}}\,\,-\,\,D\,a^2\,n^2\,\,+\,\,{\cal O}\Lb n^3\Rb\,\,
=\,\, \omega_{\mbox{\tiny BFKL}}\,\,-\,\,D\,\beta^2
\eeq
where $\omega_{\mbox{\tiny BFKL}}\,=\,4\,\ln2\, \bas$; $ D\,=\,14\, \zeta(3) \,\bas$.

Therefore in this approximation the Green function takes the form
\bea \label{GRFUD}
G\Lb Y, \kappa_{fin} | 0, \kappa_{in}\Rb \,\,&=&\,\,e^{\omega_{\mbox{\tiny BFKL}}\,Y}\,\sum_{n=0}^{\infty} \phi_n\Lb \kappa_{in},\beta\Rb\, \phi_n\Lb \kappa_{fin},\beta\Rb\,e^{ - D\,Y\,a^2 \,n^2}\nn\\
 &\to&\,\,e^{\omega_{\mbox{\tiny BFKL}}\,Y}\,\int^{\infty}_0 \,d \beta \,\phi_0\Lb \kappa_{in},\beta\Rb\, \phi_0\Lb \kappa_{fin}, \beta\Rb\,e^{ - D\,Y\,\beta^2}
 \eea
 The main contribution proportional to $e^{\omega_{\mbox{\tiny BFKL}}\,Y}$ stems from small $\beta$'s where we can use 
   \eq{EVND}. Taking the integral over $\beta$ in \eq{GRFUD} we obtain the following Green's function at large values of $Y$:
\bea \label{GFUDA}
&&G\Lb Y, \kappa_{fin} | 0, \kappa_{in}\Rb \,\,=\\
&&\,\,\,\,\frac{1}{\sqrt{\Lb \kappa_{fin}\,+\,4\Rb\,\Lb\kappa_{in}+4\Rb}} 
\,\h\,e^{\omega_{\mbox{\tiny BFKL}}Y}\,\sqrt{\frac{\pi}{D\,Y}}\Bigg\{ e^{ -  \frac{\Lb L\Lb \kappa_{fin}\Rb \,-\,L\Lb \kappa_{in}\Rb\Rb^2}{4\,D\,a^2\,Y}}\,\,-\,\,e^{ -  \frac{\Lb L\Lb \kappa_{fin}\Rb \,+\,L\Lb \kappa_{in}\Rb\,\,+\,\,2\,b_\phi\Rb^2}{4\,D\,a^2\,Y} }\Bigg\}\nn
 \eea
   
   One can see that at large $Y$ Green function $G\Lb Y, \kappa_{fin} | 0, \kappa_{in}\Rb\,\,\propto\,\,\Lb D \,Y\Rb^{-3/2}\,e^{\omega_{\mbox{\tiny BFKL}}Y}$, which should be compared with  the massless BFKL case for which $G\Lb Y, \kappa_{fin} | 0, \kappa_{in}\Rb\,\,\propto\,\,\Lb D \,Y\Rb^{-1/2}\,e^{\omega_{\mbox{\tiny BFKL}}Y}$. 
   It is related to the fact, that in the massive case the diffusion approximation is valid only at large positive $\kappa$ with a boundary condition at fixed $\kappa$.
     
\subsubsection{Evolution method}

In this method the leading  $\omega$ - plane singularity is extracted using an evolution in rapidity
$Y$,
\begin{equation}\label{NS6}
\frac{\partial \Psi}{\partial Y}\,\,=\,\,\bas\,\int d^{2}k'\, K\left(k,k'\right)\Psi\left(k',Y\right),
\end{equation}
where 
\beq \label{NS7}
\Psi\left(k,Y\right)\,\,=\,\,\int^{ \epsilon + i \infty}_{\epsilon - i \infty}\frac{d \om}{2 \pi i} e^{\om Y}\,\phi_\om\Lb \kappa\Rb,
\eeq
and $\kappa \,=\,k^2$.

For asymptotically large $Y$ at  any initial condition  function $\Psi\Lb Y_0, k\Rb$ may be decomposed over the eigenfunctions of the hamiltonian $\mathcal{H}$,
 \begin{equation}\label{NS8}
\Psi(Y,k)\,\,=\,\,\sum_{n}c_{n}\,\phi_{n}(k)\,e^{\omega_{n}Y},\end{equation}
As has been mentioned the spectrum is   discrete  since  on the grid we always have a cutoff at large $k_{T}$.
From naive counting for asymptotically
large $Y$ we have 
\begin{align*}
\Psi(Y, k) & \sim c_{0}\phi_{0}(k)\,e^{\omega_{0}Y}\,\left(1+\frac{c_{1}}{c_{0}}e^{-\Delta\omega Y}\right),\\
\Delta\omega & =\omega_{0}-\omega_{1}>0
\end{align*}
however in reality  the situation is more complicated since we have inhomogeneous
convergence and the limits don't commute
\bea
&&\lim_{k_{max}\to\infty}\Delta\omega \,\,=\,\,0,~~~~~~~~~~~~~~\lim_{k_{max}\to\infty,Y\to\infty}\Delta\omega\, Y \,\, =\,\,0\cdot\infty={\rm undefined}
\eea
In the case of massless BFKL equation the summation over $n$ in \eq{NS8} leads to the asymptotic behaviour at high energy which has been discussed  after  \eq{GFUDA}.

\begin{figure}
\begin{tabular}{ c c  c}
\includegraphics[scale=0.5]{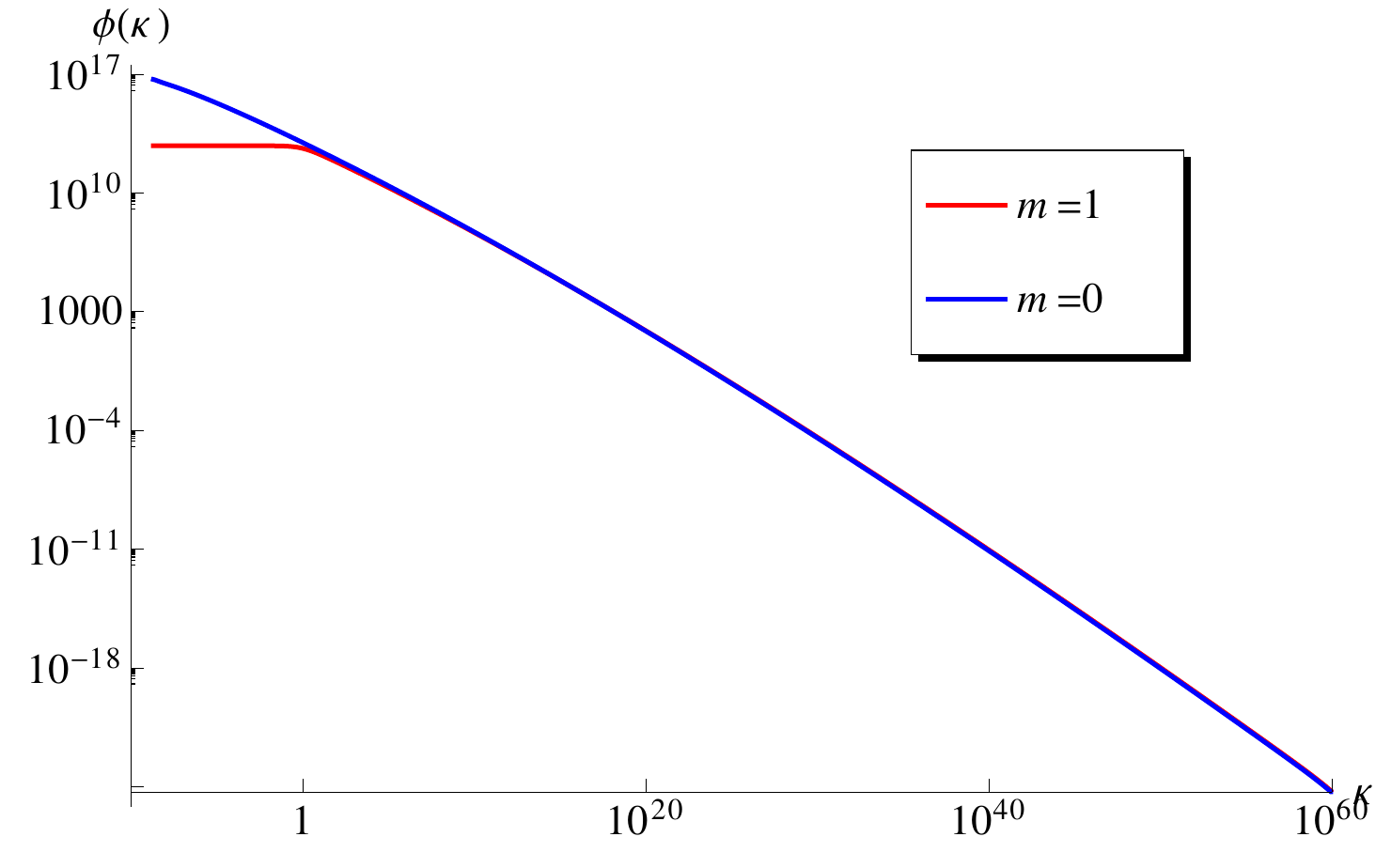}&~~~~~~~~~~&\includegraphics[scale=0.38]{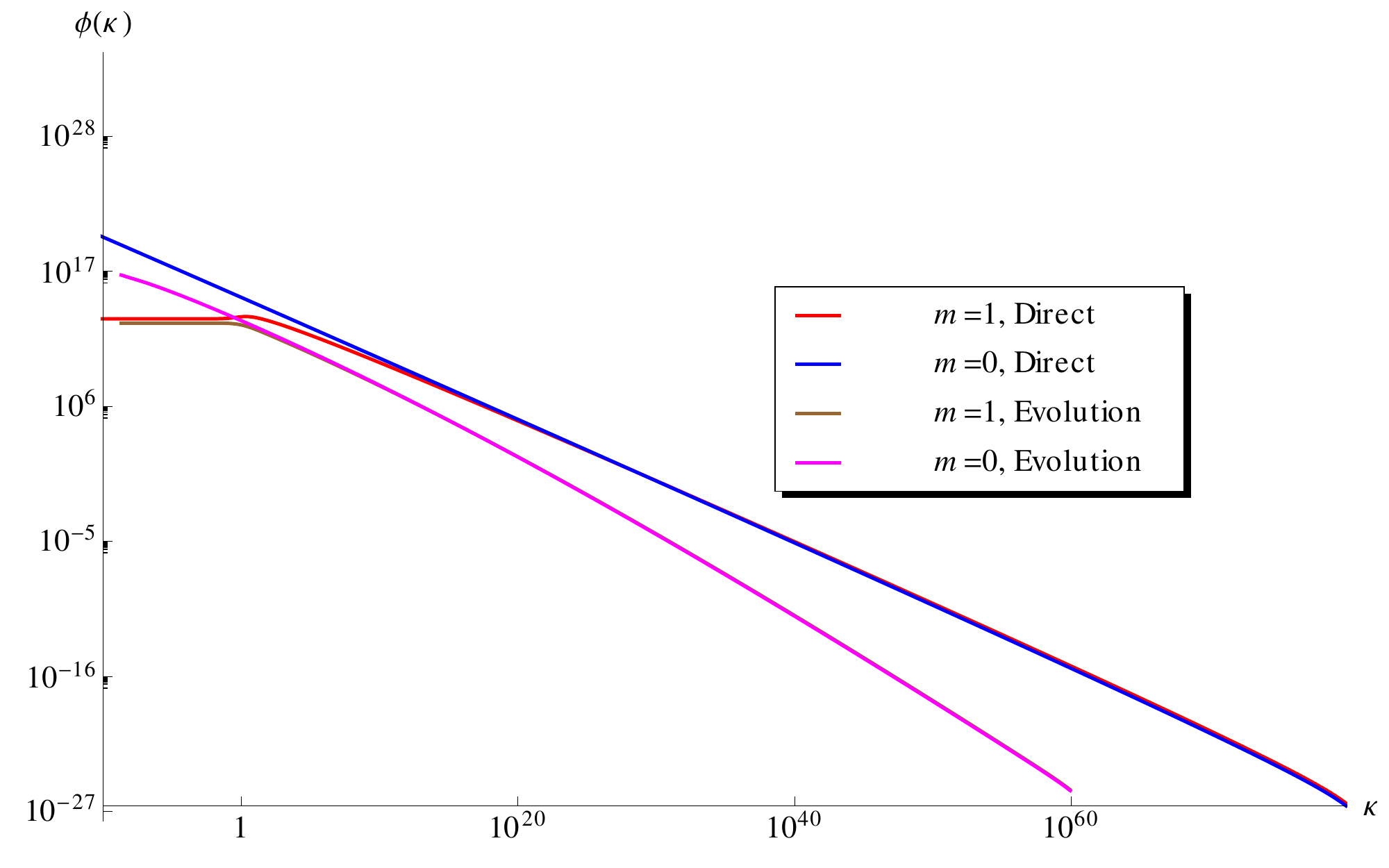}\\
\fig{evolutiontWFs}-a & &\fig{evolutiontWFs}-b\\
\end{tabular}
 \caption{\label{evolutiontWFs} Leading eigenfunctions for the massive
and massless BFKL equation extracted with evolution method (\protect\fig{evolutiontWFs}-a ). Comparison
of leading eigenfunctions extracted with direct method and evolution (\protect\fig{evolutiontWFs}-b).
Note that the eigenfunctions extracted in both methods have different
normalization, so for comparison we have multiplied the direct solution
by a normalization factor to match massive BFKL with both methods
at $k=1$.}
\end{figure}

For evolution we used a modified BK code~\cite{BKCODE} with default conditions,
$\ln\, k_{min}^{2}\in(-20,\,138)$ and $N=1024$ points in logarithmic
grid. The corresponding leading eigenvalues extracted with this method are
$\omega_{0}\,=\,0.545$ for massless case and $\omega_{0}\,=\,0.537$ for
the massive case. The results of the wave function are shown in \fig{evolutiontWFs}-a . In \fig{evolutiontWFs}-b we compare these wave functions with those extracted with direct method. For the massive case we see that both function are almost identical. For massless equation we can see that in both cases the qualitative behavior is very similar,
though quantitatively the curves differ at large $k_{T}$. Since the wave
function is suppressed there by a few orders of magnitude, we believe
that this uncertainty should not affect the physical observables.
\begin{boldmath}
\section{BFKL equation with mass at $q \neq 0$.}
\end{boldmath}
\subsection{Large impact parameter  dependence}
The kernel of the BFKL equation at $q \neq 0$ is given by \eq{K0} which we re-write using more symmetric notations for gluon momenta
\beq \label{QSY}
\vec{q}_1\,=\,\h \vec{q}\,+\,\vec{p}\,;\,\,\,\,\, \vec{q}_2\,=\,\h \vec{q}\,-\,\vec{p}\,; \,\,\,\,\,
\vec{q}\,'_1=\frac{1}{2}\vec{q}+\vec{p}\,'
\,\,\,\,\,\vec{q}^{\,\,'}_2\,=\,\h \vec{q}\,-\,\vec{p}^{\,\,'}\,;\,\,\,\,\,\vec{k}\,=\,\vec{p}\,-\,\vec{p}^{\,\,'}\,
\eeq
It takes the form
\bea \label{KERQ}
&&K\Lb \vec{q},\vec{p}, \vec{p}^{\,'}\Rb\,\,=\\
&&\,\,\frac{\bas}{2 \pi}\left\{\overbrace{ \frac{1}{k^2 + m^2}\,\Lb \frac{\Lb \h \vec{q} + \vec{p}\Rb^2 + m^2}{\Lb \h \vec{q} + \vec{p}^{\,\,'}\Rb^2 + m^2}\,\,+\,\,\frac{\Lb \h \vec{q} - \vec{p}\Rb^2 + m^2}{\Lb \h \vec{q} - \vec{p}^{\,\,'}\Rb^2 + m^2}\Rb}^{\rm emission\,\,\,\, kernel: \,K_{em}\Lb \vec{q},\vec{p}, \vec{p}^{\,'}\Rb}\,\,-\,\,\underbrace{\frac{q^2 + \frac{N^2_c + 1}{N^2_c}\,m^2}{\Lb\,\Lb \h \vec{q} + \vec{p}^{\,\,'}\Rb^2 + m^2\Rb\,\Lb\,\Lb \h \vec{q} -  \vec{p}^{\,\,'}\Rb^2 + m^2\Rb}}_{\rm contact\,\,\,\,\,\,\, term: \,\,K_{ct}\Lb \vec{q},\vec{p}^{\,'}\Rb}\right\}\nn
\eea
First, we re-write this kernel in the impact parameter representation using the following formulae:
\beq \label{IP1}
K\Lb \vec{b}, \vec{p},\vec{p}^{\,\,'}\Rb\,\,=\,\,\int d^2 q e^{ i \vec{q}\cdot\vec{b}}\,K\Lb \vec{q},\vec{p}, \vec{p}^{\,'}\Rb
\eeq
\bea \label{IP2}
&&\int d^2 q e^{ i \vec{q}\cdot\vec{b}} \frac{\Lb \h \vec{q} + \vec{p}\Rb^2 + m^2}{\Lb \h \vec{q} + \vec{p}^{\,\,'}\Rb^2 + m^2}\,=\,
 e^{2 i \vec{p}\cdot \vec{b}} \Big( - \frac{1}{4} \nabla^2_b + m^2\Big) 4  \int d^2 l e^{ i \vec{l}\cdot\vec{b}}\frac{1}{(\vec{l} + \vec{k})^2 + m^2}\,\,\\
 &&=\,\,4\,e^{2 i \vec{p}\cdot \vec{b}} \Big( - \frac{1}{4} \nabla^2_b + m^2\Big) e^{2 i \vec{k}\cdot \vec{b}}\,K_0\Lb 2 m b\Rb\,=\,
 e^{2 i \vec{p}^{\,\,'}\cdot \vec{b}}\,\left\{ k^2 K_0\Lb 2 m b\Rb\,\,+\,\,2 i m \frac{\vec{k} \cdot \vec{b}}{b}\,K_1\Lb 2 m b\Rb\right\}\nn
 \eea
 where $K_i(z)$ are the modified Bessel functions of the second kind.

Using \eq{IP2} we can re-write the first two terms of \eq{KERQ} (emission kernel $K_{em}\Lb \vec{q},\vec{p}, \vec{p}^{\,'}\Rb$) in the following form
\beq \label{IP3}
K_{em}\Lb \vec{b},\vec{p}, \vec{p}^{\,'}\Rb\,\,=\,\,\bas \Lb \frac{k^2}{k^2 + m^2}\cos\Lb 2  \vec{p}^{\,\,'}\cdot \vec{b}\Rb\,\,K_0\Lb 2 m b\Rb \,\,\,+\,\,2 m \frac{\vec{k} \cdot \vec{b}}{b}\sin\Lb 2  \vec{p}^{\,\,'}\cdot \vec{b}\Rb\,\,K_1\Lb 2 m b\Rb \Rb
\eeq
In the contact term of \eq{KERQ}  we can replace $\vec{q}^{\,\,'}_2\,=\,\h \vec{q}\,-\,\vec{p}^{\,\,'}$ and obtain the following expression 
\bea \label{IP4}
K_{ct}\Lb \vec{b},\vec{q}^{\,'}_2\Rb\,\,&=&\,\,\frac{\bas}{q'^2_2 + m^2}\Lb - \nabla^2_b +  \frac{N^2_c + 1}{N^2_c}\,m^2\Rb\, e^{\vec{q}^{\,\,'}_2\cdot \vec{b}}\,K_0\Lb m b\Rb\\
&=&\,\frac{\bas}{q'^2_2 + m^2}e^{i \vec{q}^{\,\,'}_2\cdot \vec{b}} \Lb q'^2_2 K_0\Lb m b\Rb\,\,+\frac{m^2}{N^2_c}K_0\Lb m b \Rb\,+\,2 i m \frac{ \vec{q}^{\,\,'}_2 \cdot \vec{b}}{b} \,K_1\Lb m b \Rb\Rb\nn
\eea
It is worthwhile mentioning that we can replace $\vec{q}_2^{\,\,'}$ by $\vec{p}^{\,\,'}$ in this part of the kernel since we have the integration  over $q'_2$.

The part of the BFKL kernel that is responsible for the gluon reggeization for $q\neq 0$ takes the following form (see \eq{EQ}, \eq{GTR} and \eq{QSY})
\beq \label{GTRQ}
K_{reg}\Lb \vec{q},\vec{p}\Rb\,\,=\,\,\om\Lb|\h \vec{q} +\vec{p}\,|\Rb\,\,+\,\,\om\Lb|\h \vec{q} -\vec{p}\,|\Rb
\eeq
Using \eq{GTR} and \eq{IP2}  $K_{reg}\Lb \vec{q},\vec{p}\Rb$ in $b$ representation takes the form:
\beq   \label{IP5}
K_{reg}\Lb \vec{b},\vec{p}\Rb\,\,=\,\,\bas \,m^2 \cos\Lb 2\, \vec{p} \cdot \vec{b}\Rb\,\Big\{ K^2_0\Lb 2 m b\Rb\,\,+\,\,2  K^2_1\Lb 2 m b \Rb\Big\}
\eeq
Finally, the entire kernel in $b$ representation looks as follows
\beq \label{FINKER}
{\cal K}\Lb \vec{b}, \vec{p},\vec{p}^{\,\,'}\Rb\,\,=\,\,K_{em}\Lb \vec{b}, \vec{p},\vec{p}^{\,\,'}\Rb\,\,+\,\,K_{ct} \Lb \vec{b}, \vec{p}^{\,\,'}\Rb\,\,+\,\,K_{reg}\Lb \vec{b},\vec{p}\Rb\,\delta^{(2)}\Lb \vec{p} \,-\,\vec{p}^{\,\,'}\Rb
\eeq
and the massive BFKL equation takes the form
\beq \label{BFKLMB}
\frac{\partial  f \Lb \vec{b}, \vec{p}\,| Y\Rb}{\partial Y}\,\,=\,\,\int d^2 b' \,d^2 p' \,{\cal K}\Lb \vec{b}^{\,\,'}, \vec{p},\vec{p}^{\,\,'}\Rb\, f \Lb \vec{b}\,-\,\vec{b}^{\,\,'}, \vec{p}^{\,\,'}|Y\Rb
\eeq
At large $b \,\gg\,m$  kernel ${\cal K}$ falls down exponentially, namely $ ${\cal K}$ \,\propto \exp\Lb - \,m \,b \Rb$ which leads to $ f \Lb \vec{b}, \vec{p}\,| Y\Rb\,\,\propto\,\exp\Lb -\, m\,b\Rb$. Indeed, assuming that $b' \sim 1/m$ contribute to the integral over $b'$ in \eq{BFKLMB}, we can re-write this equation in the form
\bea \label{LBEQ}
&&\frac{\partial  f \Lb \vec{b}, \vec{p}\,| Y\Rb}{\partial Y}\,\,=\\
&&\,\,\int\,d^2 p' \,\left\{\int d^2 b' {\cal K}\Lb \vec{b}^{\,\,'}, \vec{p},\vec{p}^{\,\,'}\Rb\,\right\} f \Lb \vec{b}, \vec{p}^{\,\,'}|Y\Rb\,\,+\,\,\int\,d^2 p' \,{\cal K}\Lb \vec{b}, \vec{p},\vec{p}^{\,\,'}\Rb\,\left\{ \int d^2 b' f \Lb \vec{b}^{\,\,'}, \vec{p}^{\,\,'}|Y\Rb\right\}\nn
\eea
Noticing that the largest asymptotic behaviour at large $b$ stems from $K_{ct}$ we can re-write \eq{LBEQ} in the form:
\bea \label{LBEQ1}
&&\frac{\partial  f \Lb \vec{b}, \vec{p}\,| Y\Rb}{\partial Y}\,\,=\\
&&\,\,\int\,d^2 p' \,\left\{\int d^2 b' {\cal K}\Lb \vec{b}^{\,\,'}, \vec{p},\vec{p}^{\,\,'}\Rb\,\right\} f \Lb \vec{b}, \vec{p}^{\,\,'}|Y\Rb\,\,+\,\,\bas \,e^{- m b}\int\,d^2 p' \frac{ J_0\Lb p' b\Rb}{p'^2 + m^2}\,\underbrace{\left\{ \int d^2 b' f \Lb \vec{b}^{\,\,'}, \vec{p}^{\,\,'}|Y\Rb\right\}}_{\rm solution \,\, at\,\,\, q^2=0}\nn
\eea
As we have discussed, our solution at $q^2=0$ behaves as $\Lb p^2\Rb^{-\h + i \nu}$ at large $p$ but it is constant at $p \to 0$. Integral over $p'$ in the non-homogeneous term in \eq{LBEQ1} is concentrated at small values of $p' \sim 1/b \leq m$ leading only to mild power-like dependence on $b$.
Therefore, searching solution in the form: $f \Lb \vec{b}, \vec{p}\,| Y\Rb\,=\,\exp\Lb - m\,b\Rb \tilde{f}\Lb \vec{b}, \vec{p}\,| Y\Rb$
we see that for $\tilde{f}\Lb \vec{b}, \vec{p}\,| Y\Rb$ we obtain an equation with the non-homogeneous term that only weakly  (power-like) falls at large $b$.

Hence we can conclude that at large impact parameters the solution to the BFKL equation with mass falls down as $\exp\Lb - m\,b\Rb$ as it was expected.
\begin{boldmath}
\subsection{Equation for $\langle|b^2|\rangle$}
\end{boldmath}

\begin{figure}
\centerline{\includegraphics[scale=0.5]{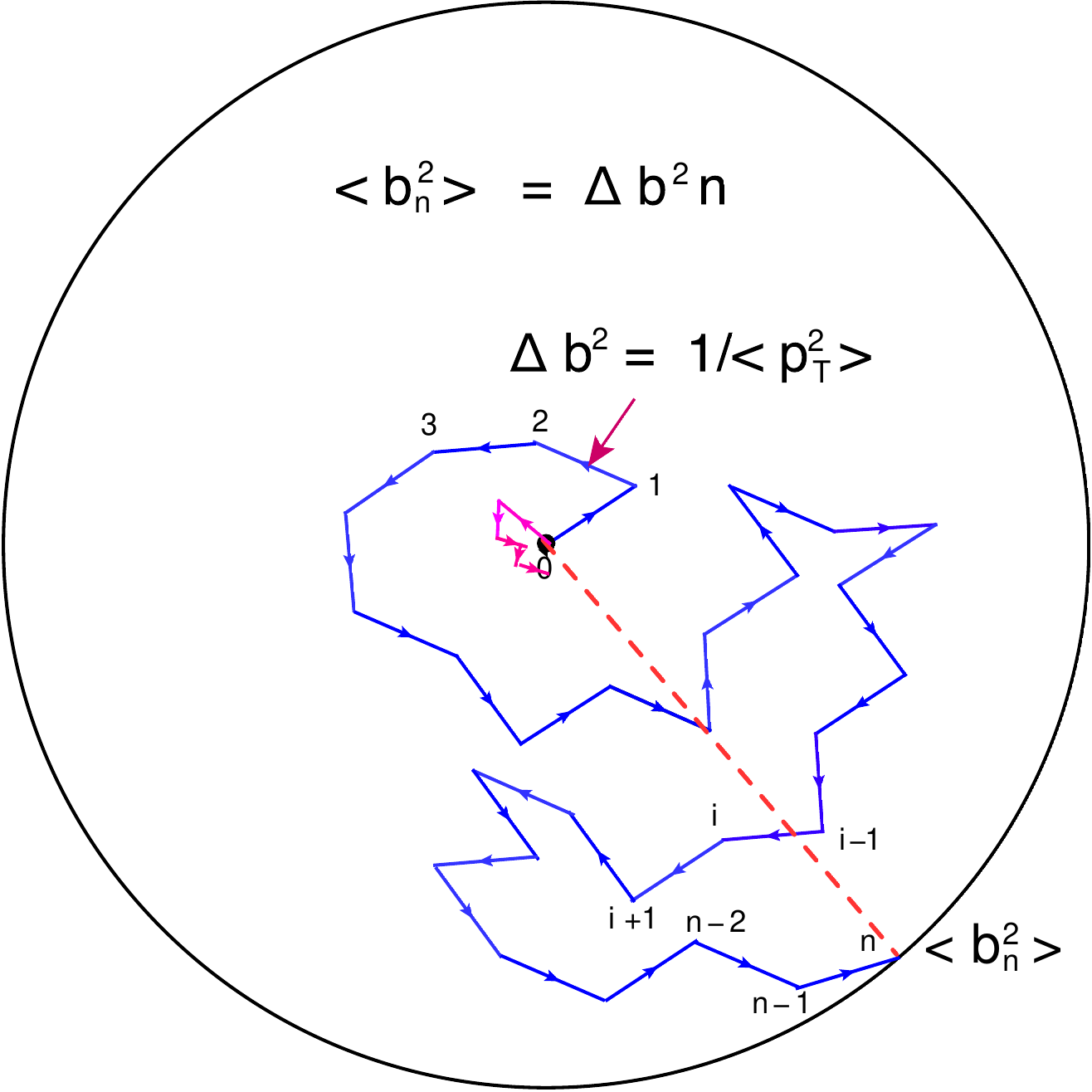}}
\caption{\label{grib} Gribov's diffusion for emissions in the parton model (blue line) and in QCD (red line).}
\end{figure}

In this section we are going to derive the equation that will allow us to calculate $\langle|b^2|\rangle$ as a function of $Y$. In the parton model this observable is proportional to the number of emissions due to Gribov's diffusion\cite{GRIB} which is sketched in \fig{grib}.

\begin{figure}
\begin{center}
\includegraphics[scale=1.1]{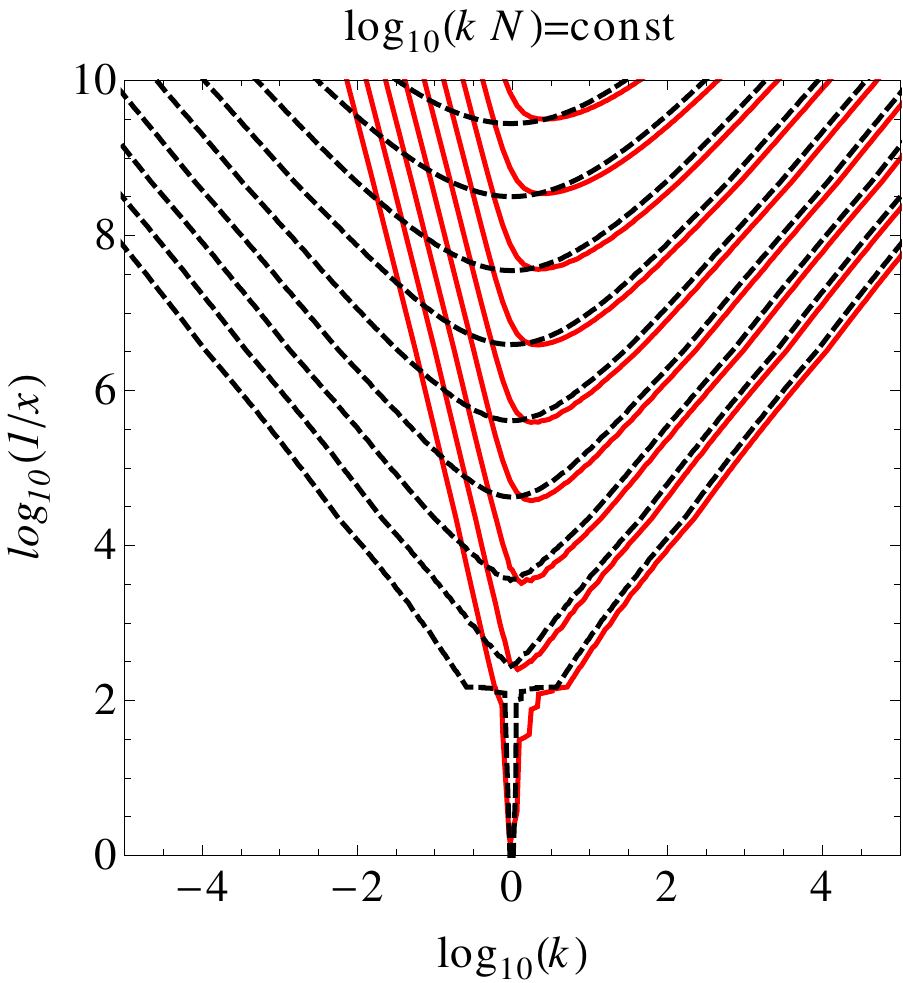}
\end{center}
 \caption{\label{contour} The contour with constant $k \Psi\Lb k, Y\Rb$ (see dotted line) for the massless BFKL equation and for the BFKL equation with mass (see solid red line) .}
  \end{figure}

The average $ b^2$ after $n$ emissions is equal
\beq \label{B21}
\langle| b^2_n |\rangle\,\,=\,\,\Delta b^2 \,n\,\,=\,\,\frac{1}{\langle| p^2_T |\rangle}\,n
\eeq
Since the average number of emissions at given $Y$ is proportional to $Y$ and the average $\langle| p^2_T |\rangle$ is a constant independent from $Y$ in the parton model,  $\langle| b^2 |\rangle\,\,=\,\,4 \,\alpha'_\pom \, Y$ where $\alpha'_\pom$ is the slope of the Pomeron trajectory.  In QCD the average transverse momentum increases with energy $Y$. We plot in \fig{contour} the  contours on which function $k \,\Psi\Lb k, Y\Rb$ (see \eq{NS6})  is constant. One can see that for the massive BFKL equation the average  $p_T$ are larger than the values of $p_T$ in initial conditions and they grow with $Y$. One can see from \eq{B21}
that $\Delta b^2 $ decreases at large $Y$ leading to $\langle| b^2_n |\rangle\,\xrightarrow{\mbox{ $Y\,\,\gg\,\,1$}} \,0$ since $\langle| p^2_T |\rangle$ increases faster than $Y$ ( see \fig{contour}). Therefore, we expect that in QCD $\langle| b^2 |\rangle $ for the massive BFKL Pomeron does not depend on $Y$ repeating the main features of the massless BFKL Pomeron.

We would like to stress that this discussion is based on the uncertainty principle  $\Delta p_T \Delta b \sim 1$. \fig{contour} shows that if we replace in \eq{B21} $1/\langle| p^2_T |\rangle$ by $ \langle| 1/ p^2_T |\rangle$ we can expect that massive BFKL   equation will lead to Gribov's diffusion since  $ \langle | 1/ p^2_T |\rangle \propto 1/m^2$.  Therefore, we need to calculate $\langle| b^2|\rangle$
for massive BFKL Pomeron to justify the simple picture that stems from \fig{grib}.

 The general expression for $\langle| b^2|\rangle$ takes the form \footnote{ \eq{B22} determines  the average $b^2$    from the imaginary part of the scattering amplitude and gives the easiest way for calculations. However, we can calculate $\langle| b^2_n |\rangle$ from the elastic cross section: viz. $ \langle| b^2_n |\rangle = \int d^2 b\,b^2 \, f^2 \Lb \vec{b}, \vec{p}\,| Y\Rb/\int d^2 b\, f^2 \Lb \vec{b}, \vec{p}\,| Y\Rb$. This definition leads to  $\langle| b^2_n |\rangle$ in two times larger than from \eq{B22}.}

 \beq \label{B22}
\langle| b^2|\rangle\,\,=\,\,\frac{\int d^2 b\,b^2 \, f \Lb \vec{b}, \vec{p}\,| Y\Rb}{ \int d^2 b\, f \Lb \vec{b}, \vec{p}\,| Y\Rb}
\eeq
and for $N\Lb  \vec{p}\,| Y\Rb\,=\,\int d^2 b\,b^2 \, f \Lb \vec{b}, \vec{p}\,| Y\Rb$ we can write the equation using the expression for the BFKL kernel in $b$ representation (see \eq{FINKER}).
However, it turns out much simpler to derive this equation using that
\beq \label{B23}
\hspace{-0.2cm}N\Lb  \vec{p}\,| Y\Rb\,=\,\int d^2 b\,b^2 \, f \Lb \vec{b}, \vec{p}\,| Y\Rb\,=\,\int d^2 b \int d^2 q  \Lb - \nabla^2_q  e^{- i \vec{q} \cdot \vec{b}}\Rb\, f\Lb \vec{q}, \vec{p},| Y\Rb\,\,=\,\,- \Lb \nabla^2_q \,\Psi\Lb \vec{q}, \vec{p} | Y\Rb\Rb|_{q = 0}
\eeq
where $\Psi$ is defined in \eq{NS6}.

Applying operator $-  \nabla^2_q$ to both parts of the evolution equation in $Y$ at $q \,\neq\,0$ we obtain
\bea \label{B24}
\hspace{-0.5cm}- \,\nabla^2_q \left\{\frac{\partial \Psi\Lb \vec{q}, \vec{p} | Y\Rb}{\partial Y}\,\right.&=&\,\left.\int d^2 p' \,{\cal K}\Lb\vec{q},\vec{p}, \vec{p}^{\,\,'}\Rb\,
\Psi\Lb \vec{q}, \vec{p}^{\,\,'} | Y\Rb\right\}\nn\\
\frac{\partial N\Lb  \vec{p} | Y\Rb}{\partial Y}&=&\int\!\!d^2 p' {\cal K}\Lb q=0,\vec{p}, \vec{p}^{\,\,'}\Rb\,
N\Lb  \vec{p}^{\,\,'} | Y\Rb +\int \!\!d^2 p' \,\Lb - \nabla^2_q {\cal K}\Lb \vec{q},\vec{p}, \vec{p}^{\,\,'}\Rb|_{q=0}\Rb \Psi\Lb \vec{p}^{\,\,'}| Y\Rb \\
 &+&\left\{ \int d^2 p' \,\Lb - \nabla_q{\cal K}\Lb \vec{q},\vec{p}, \vec{p}^{\,\,'}\Rb\Rb|_{q=0} \Lb\nabla_q \Psi\Lb \vec{q}, \vec{p}| Y\Rb\Rb|_{q=0}\right\}\,\,=\,\,0 \label{B25}
 \eea
 Using the kernel of \eq{K0}  and the notations of the momenta of gluons according to \eq{QSY} we see that $ \Lb\nabla_q{\cal K}\Lb \vec{q},\vec{p}, \vec{p}^{\,\,'}\Rb\Rb|_{q=0}\,=\,0$. 
  Using \eq{KERQ} we obtain the following expression for  $\Lb- \nabla^2_q {\cal K}\Lb \vec{q},\vec{p}, \vec{p}^{\,\,'}\Rb\Rb|_{q=0}$:
  \bea \label{KERQ1}
  &&4 {\cal K}_1\Lb p,p'\Rb\,\equiv\,\Lb- \nabla^2_q {\cal K}\Lb \vec{q},\vec{p}, \vec{p}^{\,\,'}\Rb\Rb|_{q=0}\,\,=\\
  &&\,\,\frac{2}{\,\Lb p'^2 + m^2\Rb \Lb p^2 + m^2\Rb} \,+\,\frac{4 m^2}{\Lb p'^2 + m^2\Rb^2 \Lb (\vec{p} - \vec{p}^{'})^2 + m^2\Rb} \,-\, \frac{4 m^2}{2 \Lb p'^2 + m^2\Rb \Lb p^2 + m^2\Rb\Lb (\vec{p} - \vec{p}^{'})^2 + m^2\Rb} .\nn
  \eea
  
   Hence \eq{B24} gives the equation for $N\Lb  \vec{p} | Y\Rb$.
   \bea\label{FINEQB2}
 \hspace{-0.5cm}&&\frac{\partial N\Lb  \kappa | Y\Rb}{\partial \bas  Y}\,=\\
\hspace{-0.5cm}&&\,\,\frac{\kappa +1}{\sqrt{\kappa}\sqrt{\kappa + 4  }}\ln \frac{\sqrt{\kappa + 4 } + \sqrt{\kappa}}{\sqrt{\kappa + 4 } - \sqrt{\kappa}} N\Lb \kappa|Y\Rb\,\,-\,\,\int^{\infty}_{0}\,\frac{d \kappa' N\Lb \kappa'| Y\Rb}{\sqrt{( \kappa - \kappa')^2\,+\,2 (\kappa + \kappa') + 1}}\,\,-\,\,\frac{N^2_c + 1}{2 N^2_c}\frac{1}{\kappa + 1}\int^{\infty}_0 \frac{N\Lb \kappa' | Y\Rb \,d \kappa'}{\kappa' + 1}\nn\\
&&- 4 \int^{\infty}_0 d \kappa' \Bigg\{ \frac{1 }{2(\kappa^2 + 1)^2}\,-\,\frac{2 \kappa + \kappa' + 2}{ 2(\kappa +1)^2 (\kappa'+ 1)\sqrt{(\kappa - \kappa')^2 + 2 ( \kappa + \kappa') + 1} } \Bigg\} \Psi\Lb\kappa'| Y\Rb\nn\\
&& + 2\Bigg\{\frac{\Lb 1 - 2\kappa\Rb}{\kappa\Lb \kappa + 4\Rb^2 }\,+\,
\frac{\Lb 4 + 6\kappa - \kappa^2\Rb}{2 \sqrt{\kappa} \Lb \kappa + 4\Rb \Lb \kappa + 1\Rb}\omega\Lb \kappa\Rb\Bigg\}
\Psi\Lb\kappa| Y\Rb\nn
\eea  
  Two remarks are needed: first, we substitute $N_c=3$ in the last two terms; and second, the last term stems from the expansion of the gluon trajectory( see \eq{GTR}) in the master equation (see \eq{EQ}) where their contribution takes the form: $ \omega\Lb \vec{p} - \vec{q}/2\Rb + \omega\Lb \vec{p} - \vec{q}/2\Rb$. 
  
\begin{figure}[ht]
     \begin{tabular}{c c c}
     \includegraphics[width=7.5cm]{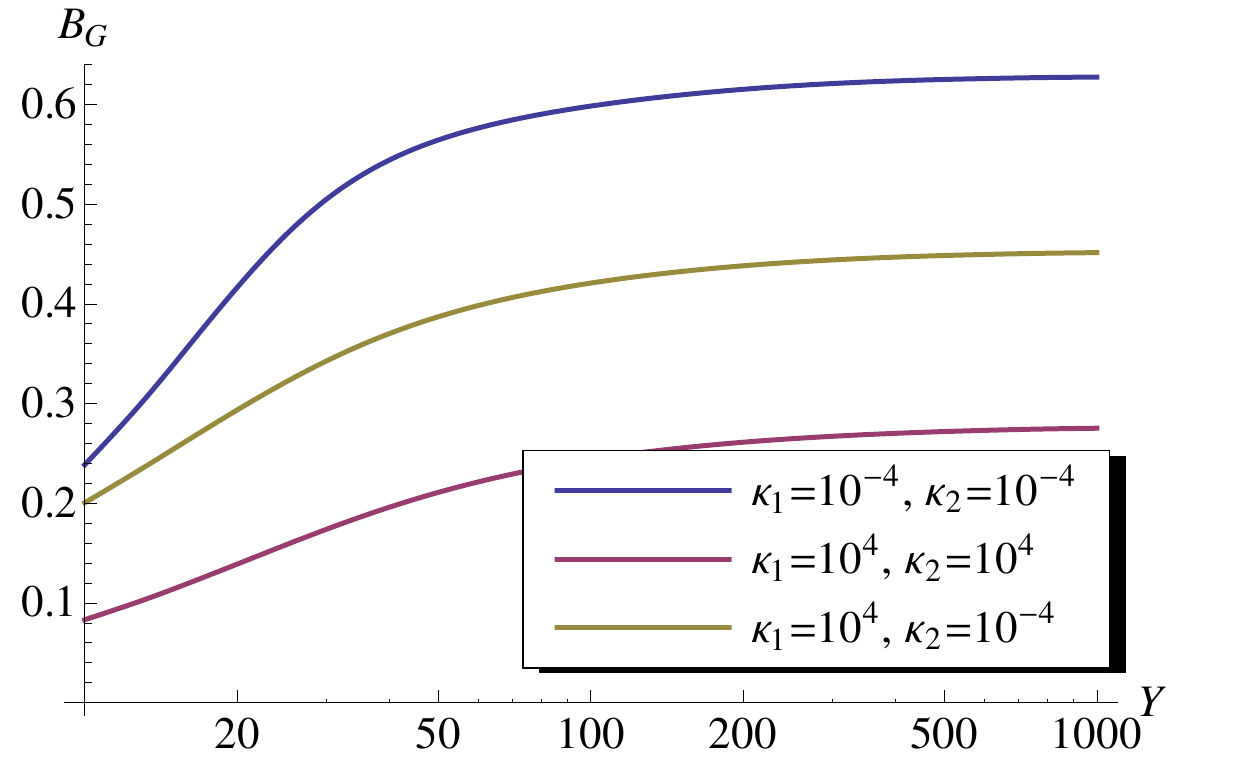}  &~~~~~~~~~~& \includegraphics[width=7.5cm]{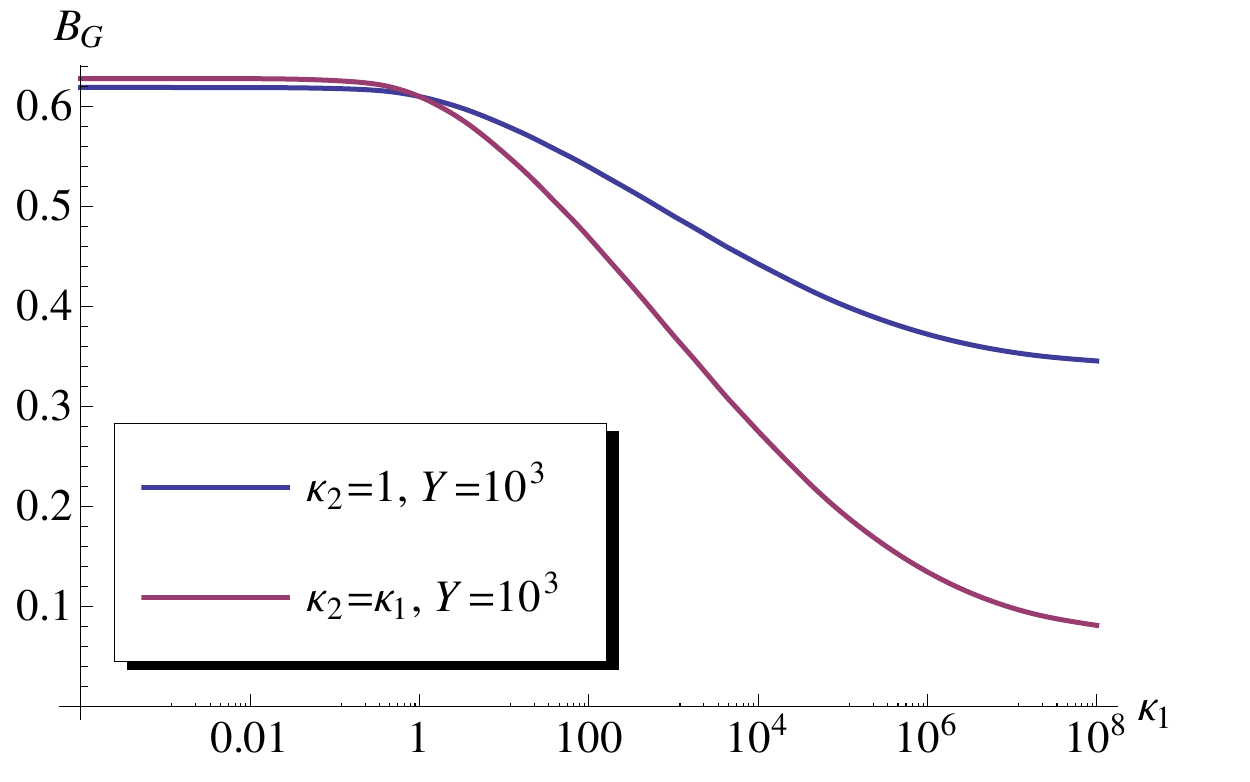} \\
     \fig{bel}-a &~~~~~~~~~~~~~~&\fig{bel}-b\\
     \end{tabular}   
      \caption{  $\langle| b^2|\rangle$ versus $Y$ (\protect\fig{bel}-a) and $\kappa_1$(\protect\fig{bel}-b). }
\label{bel}
   \end{figure}

 \fig{bel} shows $\langle| b^2|\rangle$ of \eq{B22} in which we plug in the  solution to \eq{FINEQB2}.
 We can see two general features:  $\langle| b^2|\rangle$ tends to a constant at large values of $Y$ in accordance with the qualitative  discussion (see \fig{grib}  and \fig{contour} ); and  $\langle| b^2|\rangle$ 
 does not depend on $\kappa$ for $\kappa_1 <1$ and $\kappa_2 < 1$  but falls down for $\kappa\,>\,1$.
     
  \begin{boldmath}
  \subsection{Corrections of the order of $q^2$}
  \end{boldmath}
  In this section we develop a systematic approach to the BFKL  taking into account  all corrections to the BFKL equation of the order of $q^2$. Such expansion is justified for all the eigenfunctions except those whose 
eigenvalues are in the vicinity of the point $\omega=\omega_0$. As one can see 
from~\eq{EQVE0}, near this point there is a cancellation of two leading order 
terms, so the small corrections will affect position of the pole and thus 
cannot be treated in a perturbative approach.

  Expanding the BFKL kernel of \eq{K0} we obtain 
  \begin{align}
K & =K_{0}+q^{2}K_{1},\label{QC1}\\
K_{1} & =\frac{\bas}{2\pi}\left[-\frac{1}{2\left(p'^{2}+m^{2}\right)\left(m^{2}+p^{2}\right)}+\frac{m^{2}}{2\left((\vec{p}^{\,\,'} - \vec{p})^{2}+m^{2}\right)\left(p'^{2}+m^{2}\right)\left(m^{2}+p^{2}\right)}-\frac{m^{2}}{\left((\vec{p}^{\,\,'} - \vec{p})^{2}+m^{2}\right)\left(p'^{2}+m^{2}\right)^{2}}\right. \nn\\
 & +\left.\frac{N_{c}^{2}+1}{N_{c}^{2}}\frac{m^{4}}{2\left(p'^{2}+m^{2}\right)^{3}\left(m^{2}+p^2\right)}\right]+  \label{QC2}\nn\\
 & +\frac{\bas}{2\pi}\delta^{(2)}(\vec{p} - \vec{p}^{\,\,'})\left[\frac{m^{2}\left(m^{2}-2p^{2}\right)}{p^{2}\left(p^{2}+4m^{2}\right)^{2}}+\frac{m^{2}\left(p^{4}-6m^{2}p^{2}-4m^{4}\right)\log\left(\frac{\sqrt{p^{2}+4m^{2}}+p}{\sqrt{p^{2}+4m^{2}}-p}\right)}{2p^{3}\left(p^{2}+4m^{2}\right)^{5/2}}\right]
\end{align}
where $K_0$ is the BFKL kernel at $q=0$. \eq{QC2} gives the emission part of the kernel, while \eq{QC2} stems from the reggeization term of the kernel which has a general form $\om\Lb \Lb \h \vec{q}\,-\, \vec{p} \Rb^2\Rb \,+\,\om\Lb \Lb   \h \vec{q}\,+\,\vec{p} \Rb^2\Rb$ (see \eq{GTR}). Rigorously speaking at small values of $q$ the expansion has two types  of corrections: the first contribution  is proportional to  $q^2$  and the second one which is proportional to  $  \Lb\vec{p} \cdot  \vec{q}\Rb^2$. However, below  we will assume that the wave function does not depend on orientation of the vector $q$  (this is correct assumption since conformal spin is zero for the ground state), so after integration (averaging) over the orientations of $\vec{p}$ we will get for such corrections $ \Lb\vec{p} \cdot  \vec{q}\Rb^2\,=\, \h q^2 \,p^2$. Deriving  \eq{QC2} we  performed this averaging assuming that the wave function does not depend on the orientation of vector $q$. The fact that we do not ha
 ve
  the term of the order of $\Lb\vec{p} \cdot  \vec{q}\Rb$ in the expansion of the BFKL kernel supports our assumption.

Considering $K_1 q^2$ as perturbation we obtain the following  expression for the shift of the eigenvalue of the BFKL equation 
\beq \label{QC4}
\left.\frac{d\omega_{n}}{dq^{2}}\right|_{q=0}=\frac{\int d\kappa_{1}d\kappa_{2}\phi_{n}\left(\kappa_{1}\right)\phi_{n}\left(\kappa_{2}\right)K_{1}\left(\kappa_{1},\kappa_{2}\right)}{\int d\kappa_{1}\left|\phi_{n}\left(\kappa_{1}\right)\right|^{2}}
\eeq
\begin{figure}
\begin{center}
\includegraphics[scale=0.5]{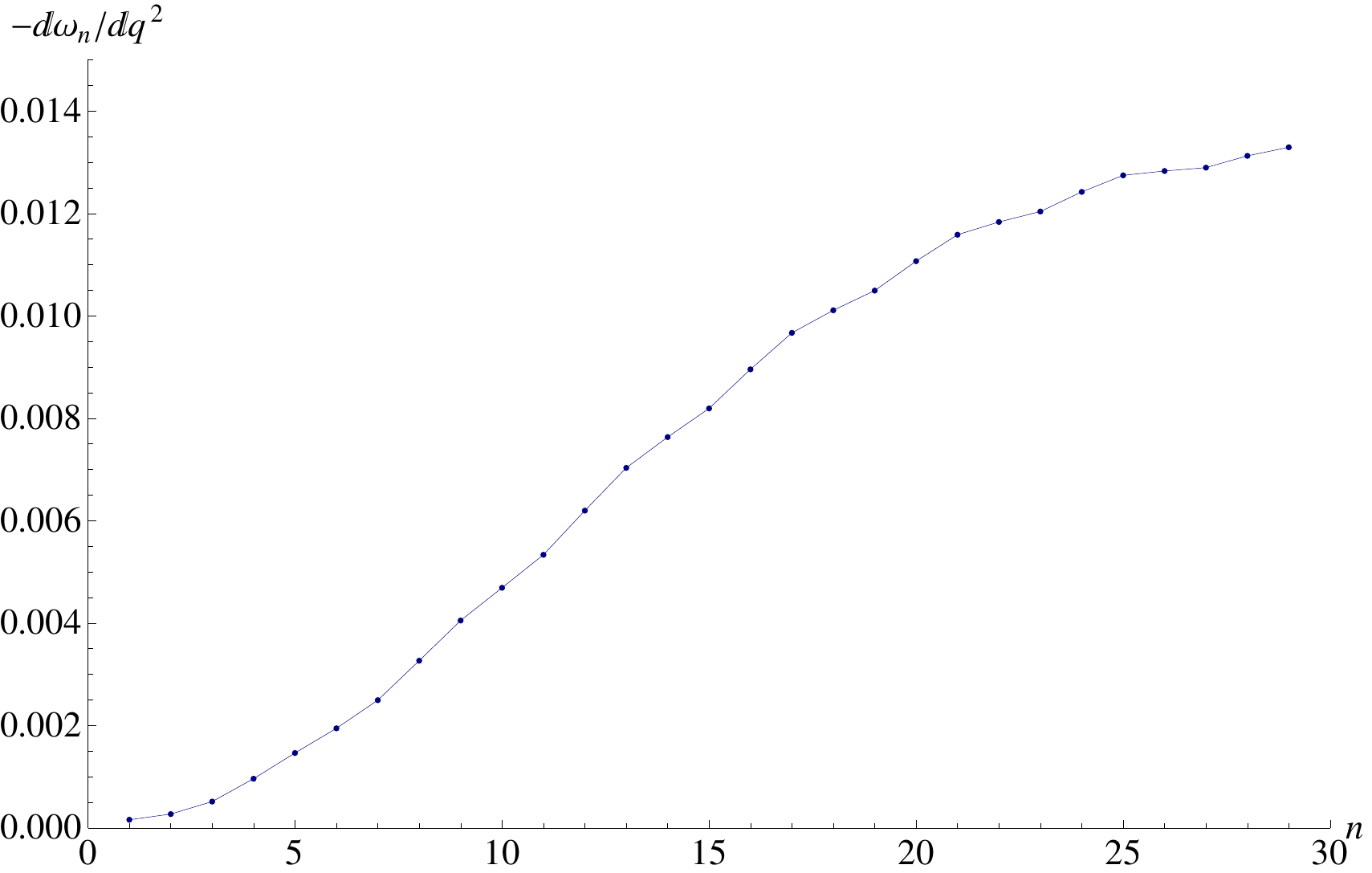}
\end{center}
 \caption{The shift in the eigenvalues $ d \omega_n/d q^2$  due to $q^2$-dependence of the BFKL kernel. $n$ is the number of roots in the eigenfunctions. }
 \label{omq}
  \end{figure}
$d\omega_{n}/dq^{2}$ is plotted in \fig{omq} as a function of $n$ where $n$ is the number of zeroes in the eigenfunction. One can see that at $n\,=\,0$  $d\omega_{n}/dq^{2}$ is equal to zero and at small $n$ it behaves as $d\omega_{n}/dq^{2}\,\,=\,\,a_q\,n^2$.

The corrections to the eigenfunctions look as follows:
\begin{equation}
\left.\frac{d\phi_{n}\left(\kappa,q\right)}{dq^{2}}\right|_{q=0}=\sum_{k\not=n}\frac{\phi_{k}(\kappa)}{\omega_{n}-\omega_{k}}\frac{\int d\kappa_{1}d\kappa_{2}\phi_{n}\left(\kappa_{1}\right)\phi_{k}\left(\kappa_{2}\right)K_{1}\left(\kappa_{1},\kappa_{2}\right)}{\int d\kappa_{1}\left|\phi_{k}\left(\kappa_{1}\right)\right|^{2}}.\label{QC5}
\end{equation}
\eq{QC4} and \eq{QC5} allows us to calculate the elastic slope of the scattering amplitude which is defined as

\beq \label{QC6}
B\Lb Y; k_{fin}\Rb\,\,=\,\,\frac{1}{4}\, \langle b^2 \rangle\,\,=\,\,2\, \frac{d\, \mbox{Im}A\Lb Y, k_{fin}| q\Rb}{ d \,q^2}{\Big|}_{q = 0}{\Bigg/}\mbox{Im}A\Lb Y, k_{fin}| q = 0\Rb
\eeq
where $A\Lb  Y, k_{fin}| q\Rb$ is the scattering amplitude which is equal to $\Psi\Lb Y, k_{fin}\Rb$ of \eq{NS8} at $q = 0$. Generally speaking this observable depends on the initial condition for the scattering amplitude at $Y=0$. However, in the diffusion approximation this dependence factorizes and can be cancelled in \eq{QC6}.

Bearing this in mind we calculate $B$ for the Pomeron Green function: viz.
\beq \label{QC7}
B_G\Lb Y; k_{fin}, k_{in}\Rb\,\,=\,\,2\, \frac{d\,G\Lb Y, k_{fin},k_{in}| q\Rb}{ d \,q^2}{\Big|}_{q = 0}{\Bigg/}G\Lb Y, k_{fin},k_{in}| q = 0\Rb
\eeq
Using the general definition of the Green function, we obtain
\beq \label{GRFUQ}
G\Lb Y, \kappa_{fin} | 0, \kappa_{in}; q\Rb \,\,=\,\,\sum_{n=0}^{\infty} \phi_n\Lb \kappa_{fin}; q\Rb\, \phi_n\Lb \kappa_{fin}; q\Rb\,e^{\omega_n\Lb q \Rb Y}
\eeq
which leads to the following expression for $B_G$:
\bea \label{QC8}
B_G\Lb Y; k_{fin}, k_{in}\Rb\,\,&=&\,\,\frac{2}{G\Lb Y, k_{fin},k_{in}| q = 0\Rb}\Bigg\{
\sum_{n=0}^{\infty} \frac{d \omega}{d q^2} \,Y\,\phi_n\Lb \kappa_{fin}; q=0\Rb\, \phi_n\Lb \kappa_{fin}; q=0\Rb\,e^{\omega_n\Lb q=0 \Rb Y} \,\\
&+&\,\sum_{n=0}^{\infty} e^{\omega_n\Lb q=0 \Rb Y}\,\Big[ \frac{d \phi_n\Lb \kappa_{fin}; q\Rb}{d q^2}{\Big|}_{q =0}\, \phi_n\Lb \kappa_{fin}; q=0\Rb\,\,+\,\,\phi_n\Lb \kappa_{fin}; q=0\Rb \frac{d\, \phi_n\Lb \kappa_{fin}; q\Rb}{d q^2}{\Big|}_{q =0}\Big] \Bigg\}\nn
\eea
The first term increases with $Y$ and gives the main  contribution at large values of $Y$. As one can see from \fig{omq} at small $n$ 
$d \omega_n(q)/d q^2\,\,=\,\,a_q\,n^2\,\,=\,\,b_q\,\beta^2$. Using this expression and the diffusion approximation of \eq{EVND} we can obtain the simple formula for the first term in \eq{QC8}:
\bea \label{QC9}
B^{(1)}_G\Lb Y; k_{fin}, k_{in}\Rb\,\,&=&\,\,\frac{2}{G\Lb Y, k_{fin},k_{in}| q = 0\Rb}\,
\sum_{n=0}^{\infty} \frac{d \omega}{d q^2} \,Y\,\phi_n\Lb \kappa_{fin}; q=0\Rb\, \phi_n\Lb \kappa_{fin}; q=0\Rb\,e^{\omega_n\Lb q=0 \Rb Y}\,\,\nn\\
&=&\,\frac{2}{G\Lb Y, k_{fin},k_{in}| q = 0\Rb}\,
\int_{0}^{\infty} d \beta b_q \,\beta^2 \,Y\,\phi\Lb \kappa_{fin},\beta; q=0\Rb\, \phi\Lb \kappa_{fin},\beta ; q=0\Rb\,e^{\omega_n\Lb q=0 \Rb Y}\,\,\nn\\
&=&\,\,- 2\,b_q\,\frac{d \ln G\Lb Y, k_{fin},k_{in}| q = 0\Rb}{d \Lb D \,Y\Rb}
\eea
We can evaluate this contribution using \eq{GFUDA}. One can see that at large $Y$ $B^{(1)}\,\to\,(3/2)b_q/D$ .
Therefore, \eq{QC8} leads to $B$ which is constant as far as $Y$ dependence is concerned in a agreement with our qualitative discussion in section 4.2.

  
  \section{Conclusions}
  
    The main goal of this paper is to find out how the correct impact parameter behaviour could affect the spectrum and the eigenfunctions of the BFKL equation. We choose the BFKL equation in the non-abilean  gauge theory with the Higgs mechanism of the mass generation as the model\,for the correct $b$ behaviour at large $b$. 
    
     We found that the massive BFKL equation for all $\omega$ larger than $\omega_0 = - \h \bas$ leads to the same eigenvalues as the massless BFKL equation, and the eigenfunctions of the massive and massless euqations coincide at large momenta. At small momenta, the massive BFKL eigenfunctions approach a constant. We suggest an approximate parametrization~\ref{PARWF} for the eigenfunction which allows us to calculate the Green's function of the massive BFKL equation.
     
     Also, we found that in contrast to massive case, there is a special point $\omega = \omega_0$ in the spectrum. The eigenfunctions in the vicinity of this point have a singularity, as one can see from a simple parametrization~\eq{EFPB0} and they are different from the massless BFKL eigenfunctions. However, we do not see how this  contribution, which falls down with energy, could contribute to the physical observables at high energy. 
     
      Hence we can state that the correct behaviour at large $b$ does not influence the main properties of the BFKL equation. This fact gives us a hope that the modification of the BFKL equation due to confinement would not affect the main equations that governs the physics at high energy ( in particular, the non-linear equations of the high density QCD). 
    
    On the other hand, the massive BFKL equation that we solved here, describes the week interaction at high energy in the case of zero Weinberg angle. We  plan to find the high energy behaviour of  the scattering amplitude in electroweak theory( see Ref. \cite{BLP}) in our future publication.

    Also, we investigated the dependence on energy for the average $<|b^2|>$ which turns out to be constant at high energy  in accordance with our expectations. In other words, we do not find that the massive BFKL Pomeron generates the slope for the Pomeron trajectory. However, it turns out that the eigenvalues with the intercepts smaller than $\omega(q^2) \,<\,\omega_L = 4 \ln 2 \bas$ have this slope, namely, 
 $d \omega(q^2)/d q^2 \neq 0$ ( see \fig{omq}). This result supports our belief that correct impact parameter behaviour does not affect the main properties of the BFKL equation as far as it concerns the scattering amplitudes at high energies.
    
  
  \section{Acknowledgements}
  
    We thank our    colleagues at UTFSM, Hamburg  and Tel Aviv universities for encouraging discussions.   One of us ( L.L)  is grateful to UTFSM for hospitality during his stay when this work was started.     Our special thanks go to D. Ross and H. Kowalsky for fruitful discussions on the massive BFKL equation.
     This research was supported by the  Fondecyt (Chile) grants 1100648  and 1120920, by the grant  RBFR-13-02-01246 and by the BSF grant 2012124.

\appendix
   \section{Appendix  }  
 As we have mentioned that from the  normalizability of function $\Psi$ the trial function of \eq{TRIF}   $\ga $ should be $\ga \geq 1/2$. Sending $a \to \infty$ we can take all integrals analytically. Indeed,
 \bea \label{A1}
 \lim_{a \to \infty}\int d \kappa |\Psi\Lb \kappa\Rb|^2\,&=&\,\int^ \infty_0 \frac{d t}{(t + a^2)^{2 \ga}}\,=\,\frac{1}{2 \ga - 1} \frac{1}{a^{4 \ga - 2}};\nn\\
  \lim_{a \to \infty}\int^\infty_0 d \kappa T\Lb \kappa\Rb |\Psi\Lb \kappa\Rb |^2\,&=&\, a^{2 - 4 \ga}\Big( \frac{\ln a^2}{2 \ga - 1} \,+\,\int^\infty_0 \frac{ d t}{(t + 1)^{2 \ga}} \ln t\Big);\\
  \lim_{a \to \infty}\int^\infty_0 d \kappa \int^\infty_0 d \kappa' \frac{|\Psi\Lb \kappa\Rb|^2}{ \sqrt{ \kappa - \kappa)^2 + 2 ( \kappa + \kappa') + 1}} \,&=&\,2 \int^\infty_0 \frac{d t}{(t + a^2)^{2 \ga}}\Big( \ln \frac{t + a^2}{\sqrt{t}}\,+\,\int^\infty_0\frac{r d y \ln y}{(y+1)^{\ga + 1}}\Big)\nn
  \eea   
  
  Hence the energy   is equal to the following expression with this rial function
  \beq \label{A2}
  E_{a \to \infty}\,\,=\,\,2 \Lb 2 \
  ga - 1\Rb \int^\infty_0 \frac{d t}{( t + 1)^{2 \ga}}\,\ln \frac{t}{t + 1}\,\,-\,\,2 \ga \int^\infty_0 \frac{d t}{( t + 1)^{\ga + 1}}\,\ln t
  \eeq
  
  \eq{A2} can be re-written in a different form, viz.
  \bea \label{A3}
&&  E_{a \to \infty}\,=\,2 ( 2 \ga -1)\Big( - \frac{1}{( 2 \ga - 1)^2} \,+\,\int_L \frac{ d z}{(z + 1)^{2 \ga}}\frac{ \ln^2(-z)}{ - 4 \pi i }\Big) \,-\,2 \ga \int_L \frac{d z}{( z + 1)^{\ga + 1}}\,\frac{ \ln^2(-z)}{ - 4 \pi i }\Big) \nn\\
&&= \, 2 ( 2 \ga -1)\Big( - \frac{1}{( 2 \ga - 1)^2} \,+\,\int^{-1}_{- \infty}  \frac{ \sin 2 \pi \ga \,d z}{(-z - 1)^{2 \ga}}\frac{ \ln^2(-z)}{ 2 \pi }\Big) \,-\,2 \ga \int^{-1}_{-\infty} \frac{\sin (\ga + 1)z)d z}{( -z - 1)^{\ga + 1}}\,\frac{ \ln^2(-z)}{ 2 \pi i }\Big)\nn\\
&&=\, 2 ( 2 \ga -1)\Big( - \frac{1}{( 2 \ga - 1)^2} \,+\,\int^{1}_{0}  \frac{ sin 2 \pi \ga \,t^{2 \ga - 2} d t}{( 1 - t)^{2 \ga}}\frac{ \ln^2 t}{ 2 \pi }\Big) \,-\,2 \ga \int^{1}_{0} \frac{\sin(\ga + 1) \pi t^{\ga - 1} d t}{( 1 - t)^{\ga + 1}}\,\frac{ \ln^2 t}{ 2 \pi  }\Big)\nn\\
&&=\,\,2 \Lb \psi\Lb \ga \Rb\,-\,\psi\Lb 2 \ga \Rb\Rb
\eea

\begin{figure}[ht]
     \begin{tabular}{c c c}
     \includegraphics[width=6.5cm]{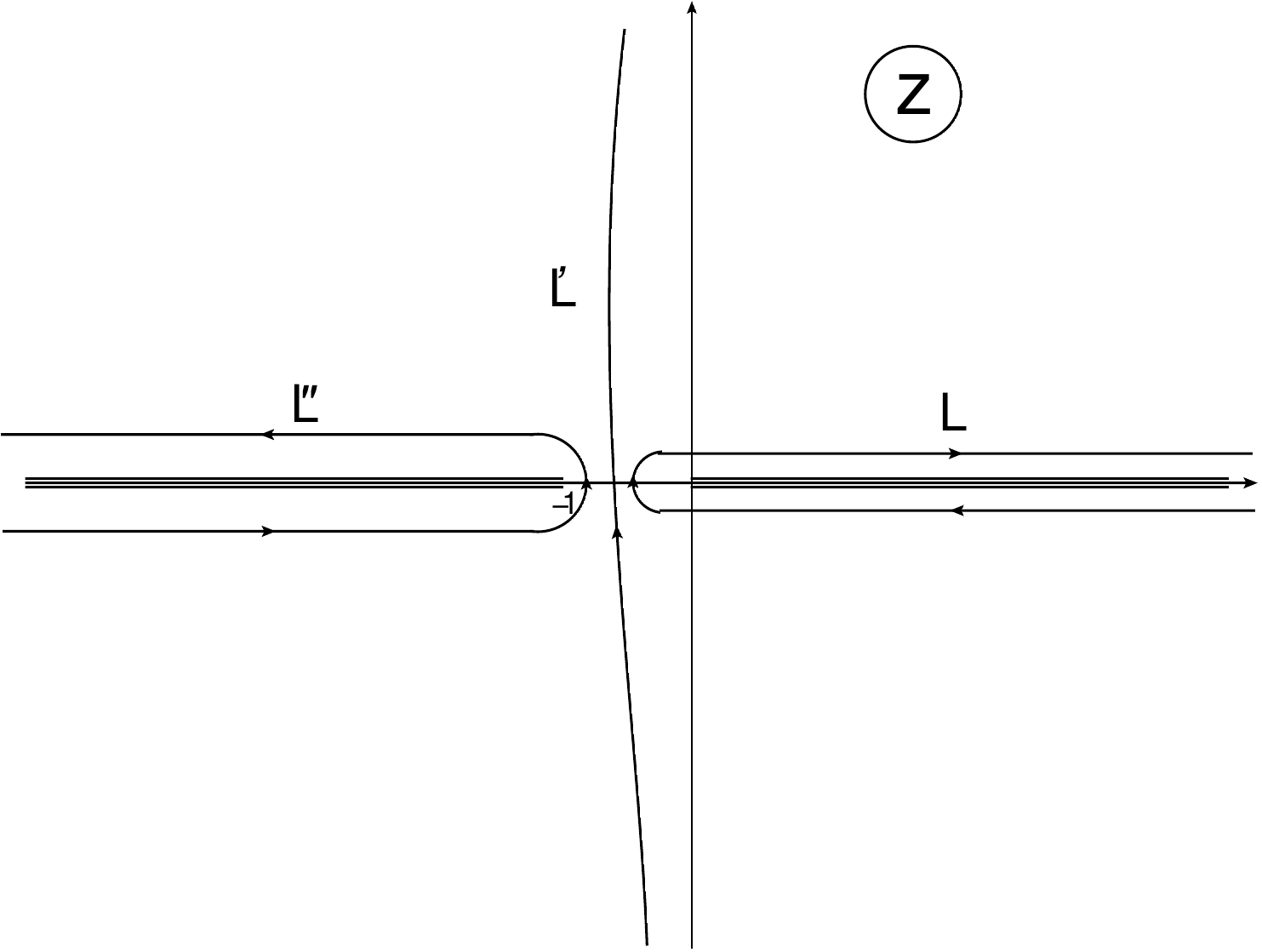}&~~~~~~~~~~~~~~  & \includegraphics[width=7.5cm]{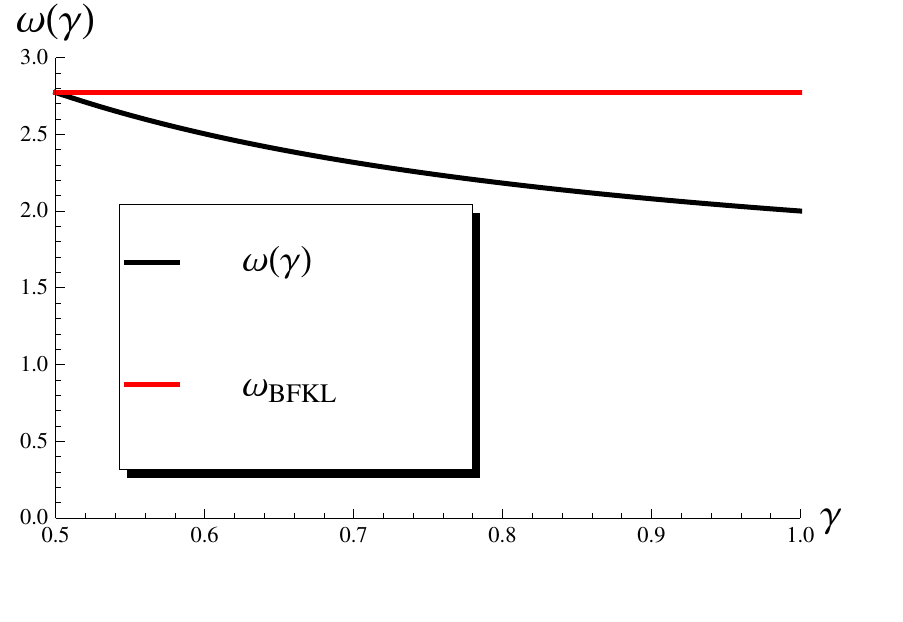} \\
     \fig{oma}-a &~~~~&\fig{oma}-b\\
     \end{tabular}   
      \caption{ The contour of integration over $z$ in \protect\eq{A3}(see \protect\fig{oma}-a) and the values of $\om = - E_{a \to \infty}$ for the analytical estimates given by \protect \eq{A3}(see \protect\fig{oma}-b). The red line shows the intercept of the BFKL Pomeron.  }
\label{oma}
   \end{figure}

 The values of $\om =  - E_{a \to \infty}$ are shown in \fig{oma}. One can see that the maximum of the intercept from the variational method is reached at $\ga = 1/2$ and it is equal to the intercept of the BFKL Pomeron.
 
 \end{document}